\documentclass[%
 reprint,
 amsmath,amssymb,
 aps,
]{revtex4-2}

\usepackage{graphicx}% Include figure files
\usepackage{dcolumn}% Align table columns on decimal point
\usepackage{bm}% bold math
\begin{document}

\preprint{APS/123-QED}

\title{Suppression of Coherent Synchrotron Radiation-Induced Emittance Growth in a Multi-Bend Deflection Line}% Force line breaks with \\

\author{Xiuji Chen}
\affiliation{Shanghai Institute of Applied Physics, Chinese Academy of Sciences, Shanghai 201800, China}
\affiliation{ShanghaiTech University, Shanghai 201210, China}

\author{Si Chen}
\email[Corresponding author]{Si Chen, chensi@zjlab.ac.cn}
\affiliation{Shanghai Advanced Research Institute, Chinese Academy of Sciences, Shanghai 201210, China}
\author{Dong Wang}
\email[Corresponding author]{Si Chen, wangdong@sinap.ac.cn}
\affiliation{Shanghai Advanced Research Institute, Chinese Academy of Sciences, Shanghai 201210, China}
\begin{abstract}
Preserving beam quality during the transportation of high-brightness electron
beams is a significant and widespread challenge in the design of modern
accelerators. The importance of this issue stems from the fact that the quality of
the beam at the accelerator's output is crucial for various applications, including
particle colliders, free-electron lasers, and synchrotron radiation sources. The
coherent synchrotron radiation (CSR) effect can degrade beam quality when a
bunch is deflected. Therefore, developing a structure that effectively suppresses
the CSR effect, especially for the short bunches is critically important. This involves protecting both the transverse emittance and
the longitudinal profile to ensure the production of a high-quality beam. In this
study, an optimization based on the reverse lattice of the beamline is proposed. This method can simplify the optimization process. Based on this approach, the Quadruple Bend
Achromat (QBA) deflection structure has been designed and optimized. Then we have derived a general solution to completely suppress the impact of steady-state CSR on the transverse plane for different topologies of QBA. Furthermore, a general condition is proposed for suppressing displacements caused by CSR in sequence drifts for isochronous structures. Simultaneously, QBA has proven to be the simplest structure that can simultaneously suppress both types of CSR effects. Simulation results for bunches with a peak current of up to $3000A$ show almost no change in transverse emittance for a large angle deflection.

\end{abstract}

\maketitle

%\tableofcontents

\section{\label{sec:level1}INTRODUCTION}
\label{sect:intro}  % \label{} allows reference to this section
Large-angle deflection of high-current ultra-short electron bunches is necessary for achieving higher luminosity in colliders and higher photon brightness in synchrotron-radiation storage rings and X-ray free electron lasers\cite{app7060607,hoffstaetter2017cbeta,article,FELDHAUS2003510,zhao2021energy,zhu2022inhibition,PhysRevAccelBeams.26.050701}. However, maintaining the quality of bunches during the deflection is of utmost importance. The presence of coherent synchrotron radiation (CSR) during deflection can result in severe degradation of bunch quality, which occurs when a number of charged particles move in a curved trajectory. In this process, the bunch emits coherent radiation that can interact with the particles in the bunch and affect their dynamics, leading to various effects on beam quality and performance.
 When a relativistic bunch passes through a dipole magnet, if the bunch length is comparable to the radiation wavelength scale, CSR radiation is generated and causes energy modulation of the bunch head and tail, leading to the distortion of the longitudinal phase space. Moreover, the coupling effect between the transverse and longitudinal directions can degrade transverse bunch quality \cite{Derbenev:291102}.

Methods for suppressing the CSR effect in compressed and uncompressed multi-bend structures have been studied extensively. The research focus on compressors is to simultaneously suppress the CSR effect and compress the long bunch length \cite{PhysRevSTAB.16.060704,PhysRevLett.110.014801,osti_1178577,PhysRevAccelBeams.20.080705,PhysRevAccelBeams.25.090701,DIMITRI2016184}. Different to the optimization objective of the compressor, the non-compressed multi-bend structures need to maintain the longitudinal profile of bunches while suppressing the CSR effect\cite{PhysRevAccelBeams.22.090701}. As non-compressed multi-bend structures are often used to bend high-intensity bunches, the intensity of the CSR effect in the isochronous arc region under the same bending angle is greater than that in the compressor. This can cause degradation of bunch length, microbunching structure, and other properties \cite{inproceedings}.

Steady-state coherent synchrotron radiation (SS CSR) in isochronous deflection structures has been extensively researched. A micro-bend is installed in the middle of the DBA structure to make $R_{56}$ close to zero\cite{Chen:IPAC2018-THPMK069,PhysRevAccelBeams.22.090701}. Simultaneously, the advance between the bends in DBA approaches $\pi$ to allow the SS CSR effects in the bends to cancel each other out\cite{PhysRevLett.110.014801}. Additionally, the micro-bend is designed with sufficiently small dimensions to make the SS CSR in the micro-bend negligible. The R-matrix method has been proposed for analyzing the displacements caused by SS CSR\cite{hajima2004r}. Then Hajima proposed the beam envelope matching method \cite{Hajima_2003}, which involves adjusting the Twiss functions to minimize the projected emittance growth. The R-matrix method was modified by Jiao\cite{PhysRevSTAB.17.060701} with the objective of analyzing the horizontal displacements in one dipole magnet, which is called the kick-point method. Related studies have demonstrated the effectiveness of this method\cite{Huang_2015,PhysRevAccelBeams.24.060701}. The integration method stands as another effective approach for predicting the emittance growth triggered by SS CSR. Its main strength lies in the independence of its optimization process from beam bunch's specific parameters, assuming the rigid beam approximation\cite{750799,PhysRevAccelBeams.19.064401}.

The variance in energy spread caused by SS CSR in a uniform magnetic field is constant when the variance in bunch length is negligible\cite{PhysRevAccelBeams.19.064401}. Based on this property, the design for these structures aimed at suppressing SS CSR is applicable to bunches with different parameters. Therefore, it is important to ensure that the variation in bunch length remains negligible across the entire beamline. Ensuring the isochronicity of the structure is essential. Simultaneously, it is necessary to ensure that the angle of the bends in the beamline is sufficiently small. A bend with a large deflection angle can significantly alter the bunch length, rendering the rigid beam approximation unreasonable. This may lead to irreversible deterioration in bunch quality, particularly in the case of ultra-short bunches. Therefore, to achieve a larger deflection angle for the bunch, a series-connected isochronous structure provides an excellent solution. Related work has been conducted in Ref\cite{PhysRevAccelBeams.25.090701}. An isochronous Triple Bend Achromat (TBA) structure with periodic solutions can achieve large angle deflection of the beam bunch while suppressing the SS CSR effect in $x$ direction. Additionally, non-SS CSR cannot be ignored either. Since it is difficult to analytically solve non-SS CSR like SS CSR, it is challenging to take it into account during the design process. A highlight work\cite{khan2021approximated} on the semi-analytical solution for non-SS CSR has made it possible to consider the entire CSR case during the design process. In this study, a new design approach has been proposed, focusing on optimization for the reverse lattice. This optimization can reduce the difficulty of design in more complex beamline scenarios. Based on this method, an isochronous Quadruple Bend Achromat (QBA) structure with periodic solutions is designed in Section.\ref{S2}. This structure can theoretically completely suppress the displacements caused by SS CSR. In Section.\ref{S3}, a general method is proposed for suppressing CSR in sequential drifts under the approximation of ultra-short bunches. Simultaneously, the necessity of QBA for suppressing all CSR effects is demonstrated. The simulation results presented in Section.IV demonstrate the effectiveness of these methods.

\section{Steady-State CSR in QBA structure}
\label{S2}

\begin{figure} 
 \begin{center}
 \includegraphics[height=1.4cm]{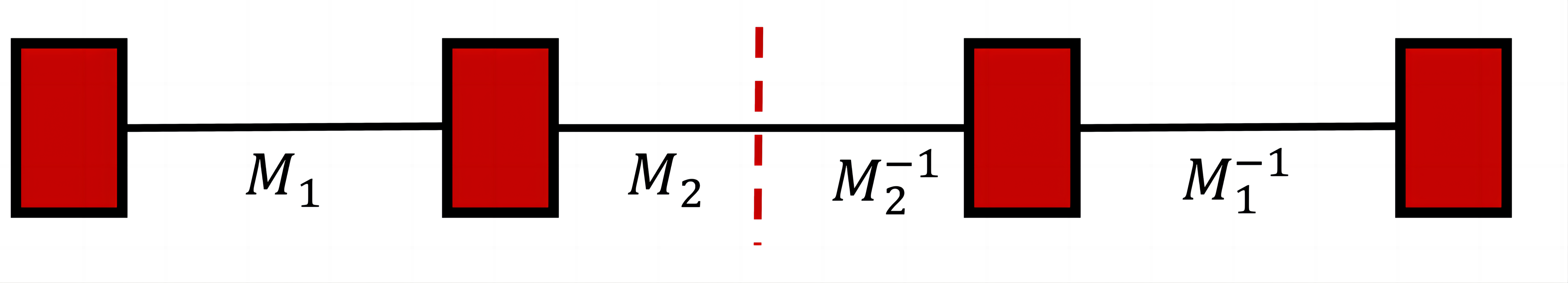}\\
 \caption{\label{QBAlayout}The layout of QBA cells. The quadrupole magnets in the matching sections are neglected in this layout. The red dashed line represents the midpoint of the QBA. $M_i^{-1}$ the inverse matrix of $M_i$}
 \end{center}
 \end{figure}

The effect of CSR on bunches is often investigated by employing a 1D projected model that neglects vertical influence and considers only longitudinal interactions. Within this approximation, the CSR effect is exclusively dependent on the natural coordinates $s$ and the longitudinal coordinate of particle $z$. As the particles are deflected in a beam transport line, the transverse displacements x and x' of the observation point $s_f$ along the beam line can be expressed as follows:
\begin{equation}
 \begin{aligned}
\hat{x}\left(s_f,z\right)&=\hat{x}_{\beta\&\eta}\left(s_f,z\right)+\hat{x}_{c s r}\left(s_f,z\right)\\
\hat{x}^{\prime}\left(s_f,z\right)&=\hat{x}_{\beta\&\eta}^{\prime}\left(s_f,z\right)+\hat{x}_{c s r}^{\prime}\left(s_f,z\right)
\end{aligned}   
\end{equation}
where $\hat{x}_{\beta\&\eta}$ and $\hat{x}’_{\beta\&\eta}$ are the oscillation motions of the particles constrained by the magnetic elements and are solely dependent on the initial coordinates of the particles in the six-dimensional phase space and lattice parameters. The transverse offsets $x_{csr}$ and $x’_{csr}$ arise owing to the coupling between the longitudinal and transverse directions, caused by the energy spread generated by the CSR effect; the energy spread induced by the CSR effect is represented by $\delta_{csr}$, at any point $s_i$, the variation in energy spread  is denoted as $\Delta \delta_{csr,i}$. In the dispersion section, the coupling terms between the transverse and longitudinal directions lead to additional transverse offsets $\Delta x_{csr,i}$ and $\Delta x’_{csr,i}$, which are produced by $\Delta \delta_{csr,i}$ generated at point $s_i$.The transverse offsets $\Delta x_{csr,i}$ and $\Delta x’_{csr,i}$ generated at observation point $s_f$ are follows :
\begin{equation}
    \begin{aligned}
&\Delta x_{c s r, i}=\Delta \delta_{c s r, i} R_{16}^{s_i \rightarrow s_f}\\
&\Delta x_{c s r, i}^{\prime}=\Delta \delta_{c s r, i} R_{26}^{s_i \rightarrow s_f}
\end{aligned}
\end{equation}
$R_{i6}^{s_i \rightarrow s_f}$ (i=1,2) represents elements $R_{16}$ and $R_{26}$ in the transfer matrix from any point $s_i$ to observation point $s_f$. The total transverse offset generated by the CSR effect from the initial point $s_0$ to the observation point $s_f$ is obtained by summing the transverse offsets $\Delta x_{csr,i}$ and $\Delta x’_{csr,i}$ produced at each point along the path. This sum can be represented in the form of an integral, given as
\begin{equation}\label{eq3}
\begin{aligned}
 \hat{x}_{\text {csr}}(z)&=\int_{s_0}^{s_f} \frac{d\delta_{\text{csr}}(z)}{d s} \cdot R_{16}^{s \rightarrow s_f} d s \\
 \hat{x’}_{\text {csr}}(z)&=\int_{s_0}^{s_f} \frac{d\delta_{\text{csr}}(z)}{d s} \cdot R_{26}^{s \rightarrow s_f} d s \\
\end{aligned}
\end{equation}
Furthermore, due to the symmetric conditions of transfer matrix, the two integrals in Eqs.~(\ref{eq3}) can be replaced by
\begin{equation}\label{eq4}
\begin{aligned}
 \hat{x}_{\text {csr}}(z)&=\int_{s_f}^{s_0} \frac{d\delta_{\text{csr}}(z)}{d s} \cdot R_{52}^{rev}(s) d s \\
 \hat{x’}_{\text {csr}}(z)&=-\int_{s_f}^{s_0} \frac{d\delta_{\text{csr}}(z)}{d s} \cdot R_{51}^{rev}(s) d s \\
\end{aligned}
\end{equation}
$R_{5i}^{rev}(s) (i=1,2)$ the elements of the transfer matrix for the reversal beamlines, and simplified as $R_{5i}$ in this study. This is done to facilitate subsequent derivations. For the symmetric structures, $R_{ij}^{rev}$ and $R_{ij}$ are identical. Therefore, $R_{ij}$ is used to represent $R_{ij}^{rev}$ in this study. This substitution is not applicable for asymmetric beamlines. Hence, the entrance of the lattice in the design work corresponds to the exit of the actual beamline for the asymmetric cases. Furthermore, only SS CSR was considered in this section. Therefore, the variance in energy spread described in Eqs.~(\ref{eq4}) satisfies
\begin{equation}\label{eq5}
\begin{aligned}
\frac{d\delta_{scr}(z)}{ds}= \begin{cases}\frac{2 N r_c}{3^{\frac{1}{3}} \rho^{\frac{2}{3}} \gamma} \int_{-\infty}^z \frac{1}{\left(z-z^{\prime}\right)^{\frac{1}{3}}} \frac{\partial \lambda\left(z^{\prime},s\right)}{\partial z^{\prime}} d z^{\prime}, & \text { if } s\in L_B \\ \\ 0, & \text { if } s \notin L_B .\end{cases}
\end{aligned}
\end{equation}
$\rho$ is the dipole radius, $\gamma$ is the relativistic factor,  $r_c$ is the classical radius of the electron,  $\lambda(z,s)$ is the normalized longitudinal density of the bunch at point $s$, N the population of the bunch. When the variance in the bunch length is neglected, the $\frac{d\delta_{csr}(s)}{ds}$ in bends is independent of $s$ under steady-state approximation. Numerous studies have demonstrated the validity of this approximation when $R_{56}$ approaches zero throughout the entire beamline\cite{PhysRevAccelBeams.24.060701,PhysRevAccelBeams.19.064401}. In order to completely cancel the SS CSR kicks, the beamline must meet certain conditions:
\begin{equation}\label{eq6}
    \begin{aligned}
I_1=\int_{s_f}^{s_0}R_{51,B}ds\rightarrow0\\
\end{aligned}
\end{equation}
\begin{equation}\label{eq61}
    \begin{aligned}
I_2=\int_{s_f}^{s_0}R_{52,B}ds\rightarrow0\\
\end{aligned}
\end{equation}
$R_{5i,B}$ the $R_{5i}$ in the bends. Therefore, to completely cancel SS CSR kicks in the beamline, the following conditions must be met: 1) isochronicity and achromaticity, 2) $I_1$ and $I_2$ must equal zero. Furthermore, to allow the deflection structure to be connected in series to achieve a larger deflection angle, an additional condition needs to be satisfied: 3) $|M_{11}+M_{12}|\leq2$. $M_{ij}$ the transfer matrix of the whole beamline. Condition 3 is known as the optically stable condition. There are no solutions that fully satisfy conditions 1)-3) for symmetric TBA structures. The minimum value of $|M_{11}+M_{22}|$ for a symmetric TBA that meets conditions 1)-2) is $17\frac{1}{4}$. For asymmetric TBA cells with three identical bends, although suitable cases exist, excessive defocusing and focusing render these solutions unusable. The common obstacle encountered in these cases is the inadequacy of available degrees of freedom to accommodate all the imposed constraints.  A simple idea is to consider symmetric Quadruple Bend Achromat (QBA) structures. There are two independent matching sections(denoted as $M_{ij,1}$ and $M_{ij,2}$ in this paper) in the QBA beamline, so it possesses the same degrees of freedom as the asymmetric TBA cells. Meanwhile, it faces fewer constraints compared to the asymmetric TBA due to its symmetric structure. The layout of the QBA was shown in Fig.~(\ref{QBAlayout}).

\begin{figure} 
  \begin{center}
  \includegraphics[height=7cm]{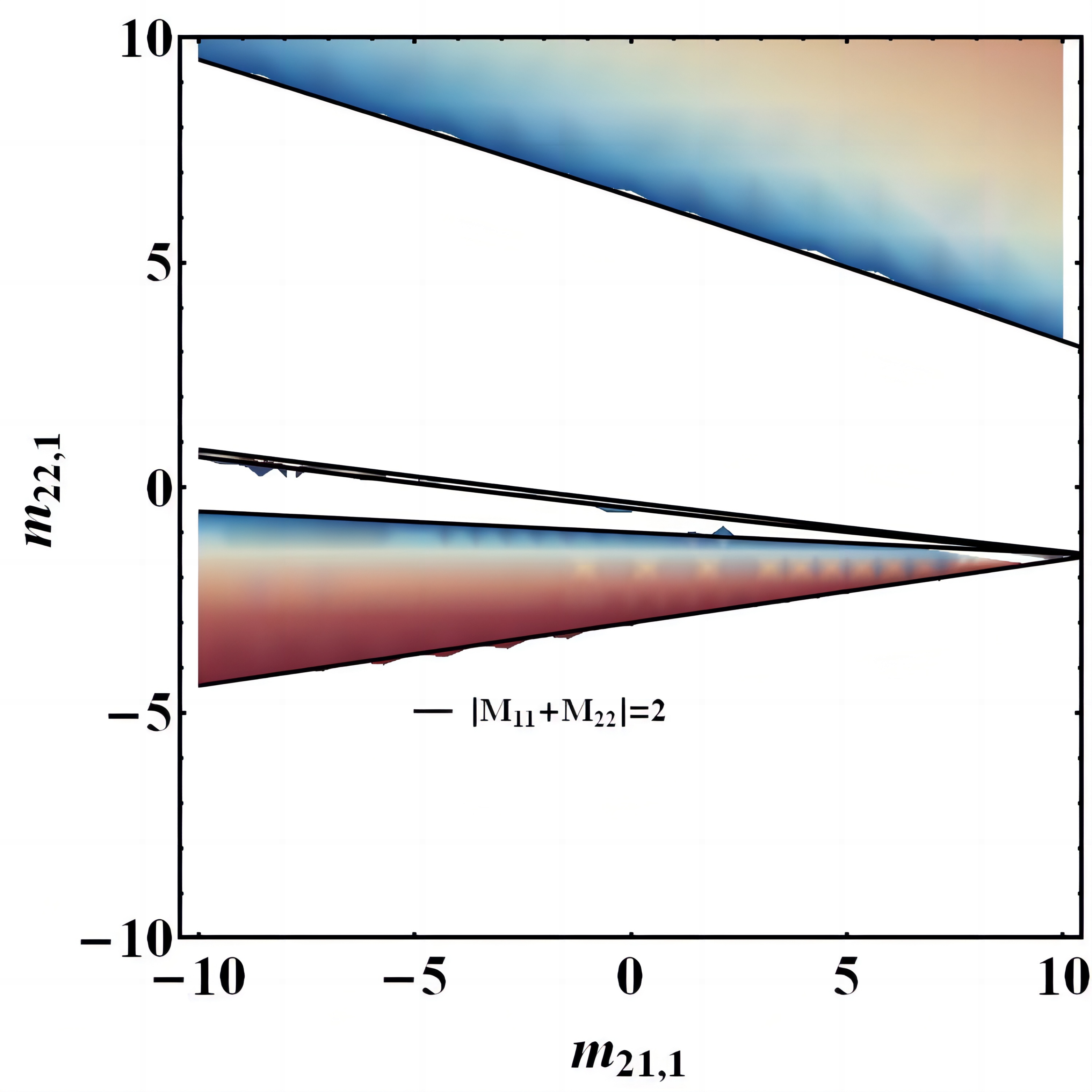} \includegraphics[height=7cm]{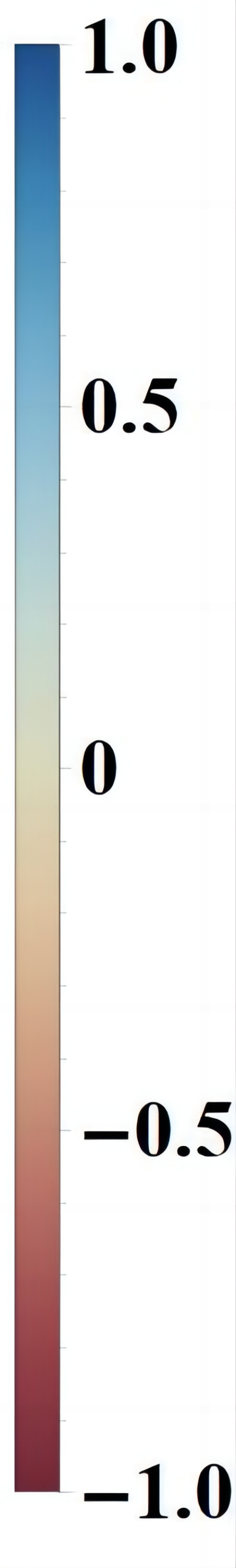}
  \caption{\label{period}The AQBA with periodic solutions.}
  \end{center}
  \end{figure}

\begin{figure} 
  \begin{center}
 \includegraphics[height=5cm]{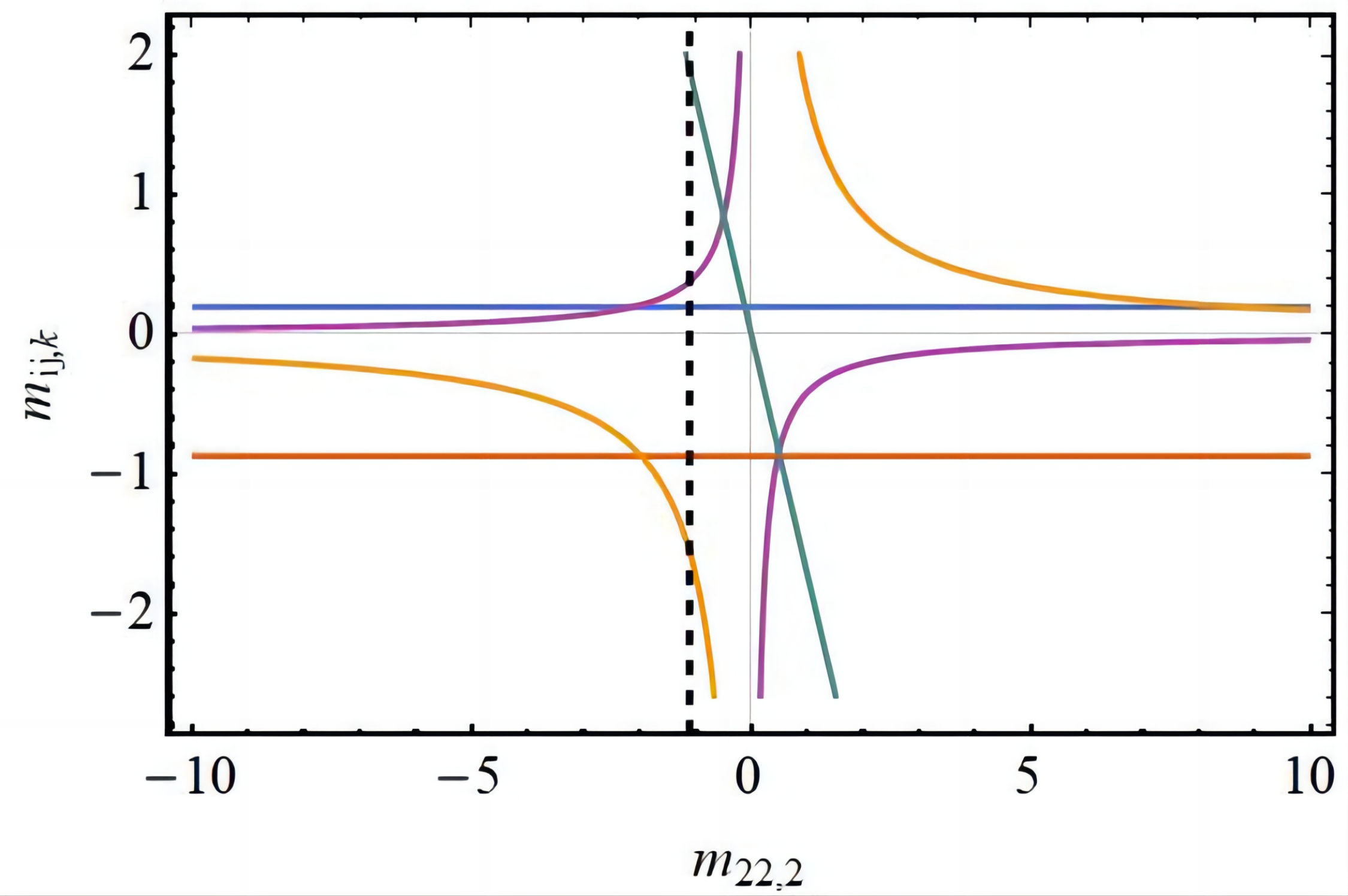}
  \caption{\label{mijk} The relationship between $m_{ij,k}$ in Eqs.~(\ref{eq8}) when $m_{21,1}$ and $m_{22,1}$ are fixed. The dash line is the work point in this study.}
  \end{center}
  \end{figure}

 In this study, only QBA with four bends of identical deflection angles was considered. There are four types of symmetric QBA structures by varying the deflection directions of the bends in the QBA.The three types are: Arc type, Chicane type, zigzag type and Dogleg type. During the derivation process, imaginary solutions were disregarded. There is a unique real solution that simultaneously sets $I_1$ and $I_2$ to zero. This indicates that the QBA has more degrees of freedom available to meet the remaining constraints. Only the Arc-like type was discussed in detail in this study(simplified as AQBA in this study). Firstly, the deflection angle of the bends needs to be small enough. This restriction ensures that the variance in bunch length can be neglected in all regions of the beamline. Otherwise, Equations (3) would be invalid. For cases involving the small-angle approximation and the ultrarelativistic approximation, the transfer matrix for the bends and the $M_i$ in both the horizontal and longitudinal planes are
 \begin{equation}\label{m1}
    \hat{M}_B=\left(\begin{array}{llll}
1 & \rho\theta & 0 & \frac{\rho\theta^2}{2} \\
0 & 1 & 0 & \theta \\
-\theta & -\frac{\rho\theta^2}{2} & 1 & -\frac{\rho\theta^3}{6} \\
0 & 0 & 0 & 1
\end{array}\right)
\end{equation}

\begin{equation}\label{a2}
    M_i=\left(\begin{array}{llll}
m_{11,i} & m_{12,i} & 0 & 0 \\
m_{21,i} & m_{22,i} & 0 & 0 \\
0 & 0 & 1 & 0 \\
0 & 0 & 0 & 1
\end{array}\right)
\end{equation}
$\theta=4^{\circ}$ the angle of the bends in AQBA. And the radius of the bends are set to $4m$. Then the solution for AQBA that satisfy conditions 1)-2) were
 \begin{equation}\label{eq8}
    \begin{aligned}
\begin{gathered}
m_{11,1}=\frac{1}{6}\left(-4-3 L_B m_{21,1}+\zeta_{++++}\right) \\ \\
m_{12,1}=-\frac{L_B\left(6+m_{22,1}(4+3 \zeta)\right)}{6 \zeta} \\ \\
m_{11,2}=-\frac{\zeta(2+3 \zeta)\left(3+5 \zeta-m_{22,1}\right)}{m_{22,2}\left(-16+3 \zeta+\zeta_{++++}\right)} \\ \\
m_{12,2}=\frac{L_B(2+3 \zeta)\left(-8+15 \zeta-3 m_{22,1}+\frac{1}{2} \zeta_{++++}\right)}{6 m_{22,2}\left(-16+3 \zeta+\zeta_{++++}\right)} \\ \\
m_{21,2}=-\frac{6 m_{22,2}(2+\zeta)}{L_B(2+3 \zeta)} \\ \\
\zeta_{++++}=\frac{12+8 m_{22,1}}{2 m_{22,1}+m_{21,1} L_B}
\end{gathered}
\end{aligned}
\end{equation}
 $L_B$ the length of bends. $\zeta=2m_{22,1}+m_{21,1}L_B$. All bends in this derivation are standard sector bends. Other types can be found in Appendix.\ref{SA}.
For the AQBA that satisfies Eqs.~(\ref{eq8}), the value of $|M_{11}+M_{22}|$ can be expressed as follows:
 \begin{widetext}
\begin{equation}\label{eq9}
    \begin{aligned}
|M_{11}+M_{22}|=\left|2\left(-\frac{1}{3}+\zeta_{++++}-\frac{4(49+27m_{22,1}-\frac{1}{2}\zeta_{++++}(41+3m_{22,1}))}{-48+9\zeta+3\zeta_{++++}}\right)\right|
\end{aligned}
\end{equation}
\end{widetext}
It means that $|M_{11}+M_{22}|$ is independent of $m_{22,2}$. Therefore, the value of $|M_{11}+M_{22}|$ can be adjusted by varying $m_{21,1}$ and $m_{22,1}$, and the $m_{ij,k}$ in Eqs.~(\ref{eq8}) can be adjusted by modifying $m_{22,2}$ when the bends are fixed. This approach will prevent over focusing or over defocusing. The area in Fig.~(\ref{period}) with "blueTored" color represents the cases where the AQBA has periodic solutions. The $m_{21,1}=-1.311$ and $m_{22,1}=0.7235$ in this study. The periodic solutions of the $\beta$ function serve as a criterion for selecting work points in Fig.~(\ref{period}). Excessively large or small $\beta$ function can couple with the unstable CSR kick, resulting in an increase in emittance. The the $m_{ij,k}$ of Eqs.~(\ref{eq8}) are shown in Fig.~(\ref{mijk}) when the $m_{21,1}$ and $m_{22,1}$ are fixed.% The parameters of the AQBA are shown in Table.\ref{tab1}. Furthermore, $I1=1e-4$ and $I2=3.4e-5$ without the small-angle approximation. 

\section{ CSR effects in the sequence drifts of QBA structure.}

Assuming that the magnet is uniform, $P'$ is the radiation point, and $P$ is the observation point, the action process of the CSR effect can be divided into the following four cases\cite{Stupakov:2002xs}: A) $P'$ is located in the preceding  drift segment, and $P$ is located in the bend. B) Both $P'$ and $P$ are located at the bend. C) $P'$ is located in the preceding drift segment, and $P$ is located in the subsequent drift segment. D) $P'$ is located in the bend, and $P$ is located in the subsequent drift segment. When the bunch length is comparable to the length of bends, there will be more scenarios to consider\cite{lou2020coherent}. A semi-analytical expression regarding the energy spread in these cases can be found in Ref\cite{khan2021approximated}. For the case A and B, the energy spread can be replaced by
\begin{equation}\label{eq11}
    \begin{aligned}
\delta_{csr,A\&B}=\frac{N r_e}{\gamma}\left(0.029 \frac{L_B \theta_T^2}{\sigma_{z}^2}+0.246 \frac{\rho^{1 / 3}\left(\theta_B-\theta_T\right)}{\sigma_{z}^{4 / 3}}\right),
\end{aligned}
\end{equation}
$\theta_{T}$ represents the angle of transition that distinguishes between the transient and steady-state regimes. $\theta_T$ is defined as $\sigma_z=1.6 \frac{\rho \theta_T^3}{24}$. This correction can be neglected when the bunch satisfies the condition $\sigma_{z}\ll L_B$. So the Eqs.~(\ref{eq11}) can be replaced by the steady-state case in this study. Furthermore, in the case of ultra-short bunches, the case C and other cases mentioned in Ref\cite{lou2020coherent} can be neglected. Then we start with the original integral form for case D:
\begin{equation}\label{eq12}
    \begin{aligned}
\frac{\delta_{csr,D}(z)}{ds}=4\frac{r_c}{\gamma}\int_{\theta}^0 \frac{\rho}{(\rho \phi+2 l)^2} \lambda\left(z-z_D\right) d \phi
\end{aligned}
\end{equation}
$\phi$ is the angular displacement in the bend of two-particle model\cite{SALDIN1997373}. $Z_D=\frac{\rho \phi^3}{24}\left(\frac{\rho \phi+4 l}{\rho \phi+l}\right)$ the slippage conditions. $r_c$ the classical electron radius. $l$ is the subsequent drift coordinate of the bends. By performing integration by parts on Eqs.~(\ref{eq12}), a more commonly used form for case D can be obtained. This integral is approximately solved by expanding the bunch distribution for small angles to the first order in Ref.\cite{khan2021approximated}. Then the variance in $\delta_{csr}(z)$ in the subsequent drift can be expressed as
\begin{equation}\label{eq12}
    \begin{aligned}
\frac{\delta_{csr,D}(z)}{ds}= \frac{4 r
_c \lambda(z)}{\gamma}\int_{\phi_0}^0 \frac{\rho}{(\rho \phi+2 l)^2} d \phi.
\end{aligned}
\end{equation}
The $\phi_0$ is the maximum displacement angle in the two-particle model, and its maximum value does not exceed $\theta$. For the case where the bunch length is much smaller than $l$, the $\phi_0$ meets the condition as follows:
\begin{equation}\label{eq12a}
    \begin{aligned}
    \sigma_z &\approx \frac{\rho \phi_0^3}{24}\left(\frac{4 l+\rho \phi_0}{l+\rho \phi_0}\right) \\ 
&\approx \frac{\rho \phi_0^3}{6}.
\end{aligned}
\end{equation}
Then the Eqs.~(\ref{eq12}) can be expressed as
\begin{equation}\label{eq12b}
    \begin{aligned}
\frac{\delta_{csr,D}(z)}{ds}= \frac{4 r
_c \lambda(z)}{\gamma\left(\frac{4l^2}{\rho \phi_0}+2l\right)}.
\end{aligned}
\end{equation}
Now let's return to Eqs.~(\ref{eq4}). After considering the csr in drift, the csr kick of $z$ at the exit of the beamline can be expressed as follows:
\begin{equation}\label{eq13}
\begin{aligned}
 \hat{x}_{\text {csr}}(z)&=\hat{x}_{\text {csr,B}}(z)+\hat{x}_{\text {csr,D}}(z)\\
 &=\int_{s_f}^{s_0} \frac{d\delta_{\text{csr,B}}}{d s} R_{52,B}+ \int_{s_f}^{s_0} \frac{d\delta_{\text{csr,D}}}{d s} R_{52,D}d s, \\
\end{aligned}
\end{equation}
\begin{equation}\label{eq15}
\begin{aligned}
 \hat{x'}_{\text {csr}}(z)&=\hat{x'}_{\text {csr,B}}(z)+\hat{x'}_{\text {csr,D}}(z)\\
 &=\int_{s_f}^{s_0} \frac{d\delta_{\text{csr,B}}}{d s} R_{51,B}+ \int_{s_f}^{s_0} \frac{d\delta_{\text{csr,D}}}{d s} R_{51,D}d s. \\
\end{aligned}
\end{equation}
The subscripts B or D represent the values of the corresponding physical quantities within the bend or drift, respectively. For a AQBA structure that satisfies the conditions of Eqs.(\ref{eq8}), the first integral on the right-hand side of Eqs.~(\ref{eq13}) and ~(\ref{eq15}) can be regarded as 0. And the $R_{5i}$ in the drift sections are constants. The CSR kick of the AQBA in this study can be replaced by
\begin{equation}\label{eq16}
\begin{aligned}
 \hat{x}_{\text {csr}}(z)=\frac{4r_c \lambda(z)}{\gamma}\sum_{i=0}^{i=3}R_{52,i} \zeta_{i} \\
\end{aligned}
\end{equation}
\begin{equation}\label{eq17}
\begin{aligned}
 \hat{x'}_{\text {csr}}(z)=-\frac{4r_c \lambda(z)}{\gamma}\sum_{i=0}^{i=3}R_{51,i} \zeta_i \\
\end{aligned}
\end{equation}

\begin{equation}\label{eq18}
\begin{aligned}
 \zeta_i&=\int_{0}^{L_{D,i}}\frac{dl}{\left(\frac{4l^2}{\rho \phi_0}+2 l\right)}\\
 &=\frac{1}{2}Log\left(\frac{2L_{D,i}}{\rho\phi_0}+1\right)
\end{aligned}
\end{equation}
The $L_{D,i}$ are the total length between the $i$th bend and the $(i+1)$th bend of re-lat. $R_{5j,i}$ is the transfer matrix for the subsequent drift of the $i$th bend of the re-lat. The integral of $\zeta_i$ is irregular, because the integrand tends to infinity when $L_{D,i}=0$. For practical physical processes, $l$ equals 0 does not fall within the scope of Case D. Therefore, similar to the integral boundaries concerning the CSR wake, the portion of the $\zeta_i$ where $l$ approaches 0 can be disregarded.  After neglecting the boundaries of $\zeta_i$, the RMS of $\delta_{csr,D}$ for a Gaussian bunch can be expressed as
\begin{equation}\label{eq19}
\begin{aligned}
\delta_{csr,D}=\sqrt{\frac{2 \sqrt{3}-3}{3 \pi}} \frac{N r_c}{\gamma \sigma_z} \log \left(\frac{2 L_{D,i}}{\rho\phi_0}+1\right).
\end{aligned}
\end{equation}
  \begin{figure} 
  \begin{center}
  \includegraphics[height=5cm]{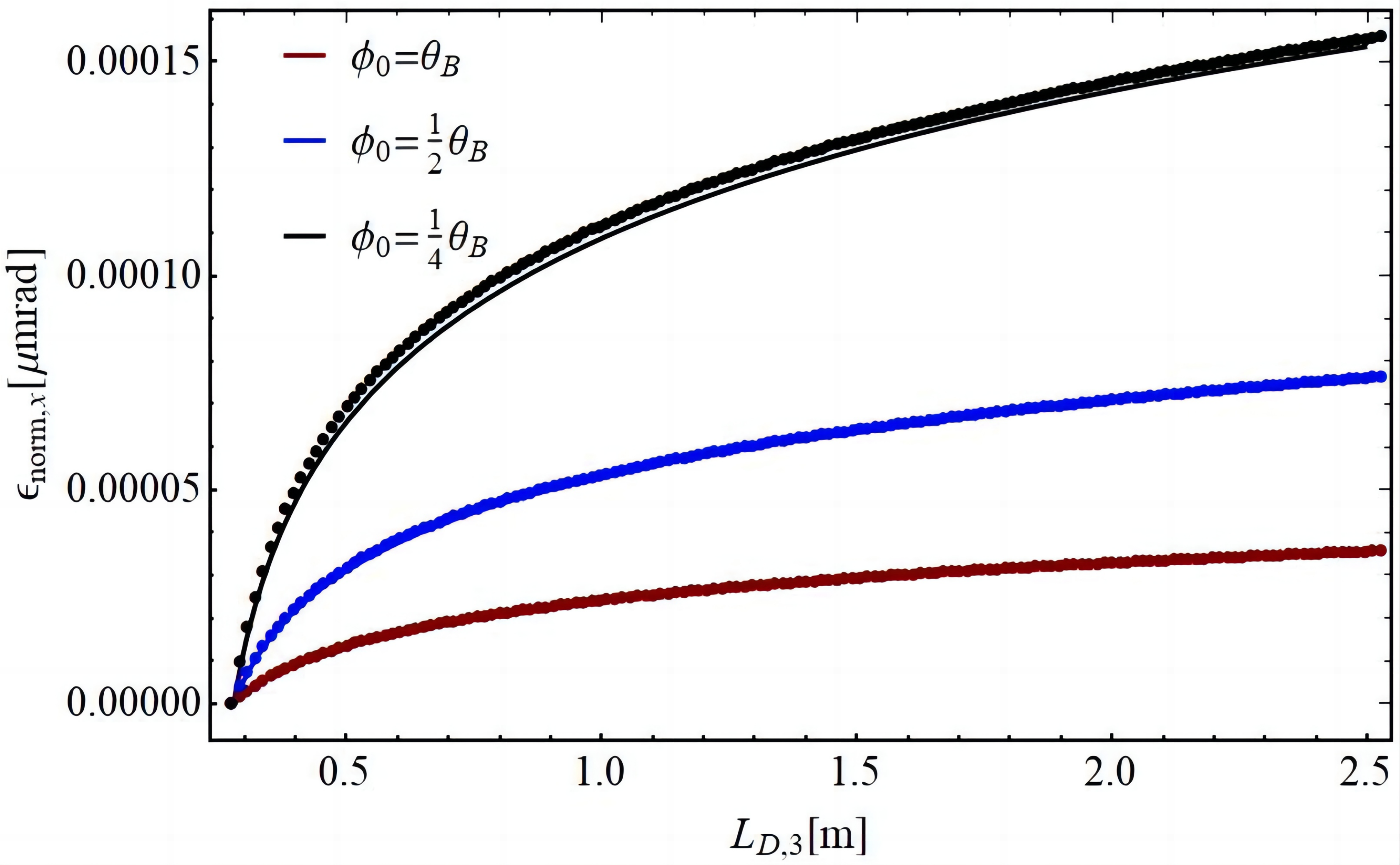}
  \caption{\label{delta_csr}The RMS energy spread caused by CSR for a Gaussian bunch with different bunch length in the sequence drift. The line: from Eqs.(\ref{eq19}). The dot: from BMAD simulation. The bunch parameters can be found in Table.\ref{beam_tab}}.
  \end{center}
  \end{figure}
This formula resembled to Eqs.(27) presented in Ref\cite{khan2021approximated}, with the sole difference being a coefficient that depends on $L_D$ and $\sigma_z$. In the case of ultra-short bunches, this coefficient tends towards 1, and disregarding it simplifies subsequent computational steps. The simulation results from BMAD\cite{SAGAN2006356} show a similar trend to Eqs.~(\ref{eq19}). The simulations for bunches of varying lengths demonstrate excellent agreement with theoretical analysis. Furthermore, Eqs.~(\ref{eq19}) can also provide an accurate prediction for the $l\sim \rho\theta_0$ region. This indicates that neglecting the boundary terms of $\zeta_i$ is reasonable, even though it is mathematically incorrect. The remaining terms of the integral provide a good prediction of the effects caused by CSR during the sequence drifts. It can be found that the CSR kick in the subsequent drift can be completely canceled for any $z$ by ensuring that the summation terms in Eqs.~(\ref{eq16}) and Eqs.~(\ref{eq17}) equal zero.  For any structure with negligible variance in bunch length, the following relationship must be satisfied for the beamline:
\begin{equation}\label{eq20}
\begin{aligned}
\prod_{i=1}^n\left(\frac{2L_{D,i}}{\rho_{i+1} \phi_{0,i+1}}\right)^{R_{51,i}}=1,\\
\end{aligned}
\end{equation}
\begin{equation}\label{eq21}
\begin{aligned}
\prod_{i=1}^n\left(\frac{2L_{D,i}}{\rho_{i+1} \phi_{0,i+1}}\right)^{R_{52,i}}=1.\\
\end{aligned}
\end{equation}
It should be noted that in the $i$-th term of the product, the parameters of $\rho_{i+1}$ and $\phi_{0,i+1}$ come from the $(i+1)$-th bend of the re-lat. Furthermore, for achromatic symmetric structures with identical bends, the sum terms of Eqs.~(\ref{eq16}) and Eqs.~(\ref{eq17}) satisfy the following relationship: 
\begin{equation}\label{eq22}
\begin{aligned}
\sum_{i=1}^{i=n}R_{52,i} \zeta_{i}=C\sum_{i=1}^{i=n}R_{51,i} 
\zeta_{i}
\end{aligned}
\end{equation}
$C$ is a parameter that is independent of  $L_{D,i}$. For a AQBA, the parameter is
\begin{equation}\label{eq222}
\begin{aligned}
C=\frac{-m_{21,2}L_B+(m_{22,1} +L_Bm_{21,1})(2m_{22,2}+L_Bm_{21,2})}{2m_{21,1}m_{22,2}+m_{21,2}(L_Bm_{21,1}-2)}.\\
\end{aligned}
\end{equation}

Therefore, there exists a set of solutions where $\hat{x}_{csr,D}$ and $\hat{x}^{\prime}_{csr_D}$ are both equal to 0. This implies that we can utilize fewer degrees of freedom to meet the conditions of Eqs.~(\ref{eq20}) and Eqs.~(\ref{eq21}) for the re-lat. This provides greater flexibility in the design process, rather than being confined to a single structure when the bends in the beamline are already fixed. For the AQBA studied in this paper which meets the conditions in Eqs.~(\ref{eq8}), the solutions are as follows:
\begin{equation}\label{eq23}
\begin{aligned}
L_{D,2}=\frac{\rho\phi_0}{2}\left(\left(1+\frac{2L_{D,1}}{\rho\phi_0}\right)^{-\frac{3\zeta+\zeta_{++++}-16}{\zeta(6+\zeta)}}-1\right)
\end{aligned}
\end{equation}
Noted the $L_{D,1}=L_{D,3}$. For the other cases presented in Appendix \ref{SA}, the solutions can be obtained by substituting $\zeta_{++++}$ with the corresponding parameters. The discussion in Appendix.\ref{SD} shows that there are no solutions that allow the TBA structures to simultaneously satisfy the conditions stated in Eqs.~(\ref{eq6})-(\ref{eq61}) and Eqs.~(\ref{eq20})-(\ref{eq21}). So, it seems that QBA is the simplest structure that can meet all the conditions.

\section{simulations for AQBA structure.}

\label{S4}

\subsection{simulations for single AQBA cells}
The parameters of the AQBA are shown in Table.\ref{tab1}. Furthermore, $I_1=1e-4$ and $I_1=3.4e-5$ without the small-angle approximation. Initially, only SS CSR was considered. Furthermore, the twiss functions for the beamline have been set to follow periodic solutions. And Then the horizontal emittance at the exit of the beamline can be expressed as\cite{PhysRevAccelBeams.19.064401}
 \begin{equation}\label{eq41}
 \begin{aligned}
 \varepsilon_x^2&=\varepsilon_{x 0}^2+\varepsilon_{x 0}\left(\sqrt{\frac{M_{12}^2}{1-M_{11}^2}} I_1^2+\sqrt{\frac{1-M_{11}^2}{M_{12}^2}}I_2^2\right)I^{2}_{csr}\\ \\
 \end{aligned}
 \end{equation}
 $\varepsilon_x$, $\varepsilon_{x_0}$ the geometric emittance at the entrance and exit of the AQBA. $I_{csr}$ is the RMS energy spread casesd by SS CSR. For the Gaussian bunch, the $I_{csr}$ is\cite{SALDIN1997373}
 \begin{equation}\label{eq42}
 \begin{aligned}
 I_{csr}= \begin{cases} 0.246 \frac{N r_e}{\gamma \rho^{2 / 3} \sigma_z^{4 / 3}} & \text { if } s\in L_B \\ \\ 0, & \text { if } s \notin L_B .\end{cases}
 \end{aligned}
 \end{equation}

  The accelerator physics simulation program BMAD\cite{SAGAN2006356} was employed for accelerator physics and CSR effect simulations. The bunch parameters was listed in Table.\ref{beam_tab}. The simulation results show that the variance in emittance is less than 0.001$\%$, and the longitudinal profile of the bunch is preserved well(not shown in this section).

  \begin{table}[!htb]
 \centering
 \caption{Parameters of AQBA in this study.}
 \label{tab1}
 \begin{tabular}{lll}
 \hline\hline Parameter &\quad\quad\quad\quad\quad  Value &\quad\quad\quad\quad\quad  Units \\
 \hline Angle &\quad\quad\quad\quad\quad  4 &\quad\quad\quad\quad\quad  $\circ$ \\
 Radius &\quad\quad\quad\quad\quad  4 &\quad\quad\quad\quad\quad  $\mathrm{m}$ \\
 $m_{11,1}$ &\quad\quad\quad\quad\quad  -0.872 &\quad\quad\quad\quad\quad  $ $ \\
 $m_{12,1}$ &\quad\quad\quad\quad\quad  0.198 &\quad\quad\quad\quad\quad $\mathrm{m}$ \\
 $m_{21,1}$ &\quad\quad\quad\quad\quad 0.724 &\quad\quad\quad\quad\quad $ \mathrm{m^{-1}}$ \\
 $m_{22,1}$ &\quad\quad\quad\quad\quad -1.311 &\quad\quad\quad\quad\quad $ $\\
 $m_{11,2}$ &\quad\quad\quad\quad\quad  -1.56 &\quad\quad\quad\quad\quad  $ $ \\
 $m_{12,2}$ &\quad\quad\quad\quad\quad  0.379 &\quad\quad\quad\quad\quad $\mathrm{m}$ \\
 $m_{21,2}$ &\quad\quad\quad\quad\quad 1.888 &\quad\quad\quad\quad\quad $ \mathrm{m^{-1}}$ \\
 $m_{22,2}$ &\quad\quad\quad\quad\quad -1.1 &\quad\quad\quad\quad\quad $ $\\
 $\zeta$ &\quad\quad\quad\quad\quad -2.42 &\quad\quad\quad\quad\quad $ $\\
 $\zeta_{++++}$ &\quad\quad\quad\quad\quad -0.6248 &\quad\quad\quad\quad\quad $ $\\
 \hline
 \hline
 \end{tabular}
 \end{table}

  \begin{table}[!htb]
 \centering
 \caption{Parameters of the beams at the entrance of AQBA.}
 \label{beam_tab}
 \begin{tabular}{lll}
 \hline\hline Parameter &\quad\quad\quad\quad\quad  Value &\quad\quad\quad\quad\quad  Units \\
 \hline Bunch charge &\quad\quad\quad\quad\quad  500 &\quad\quad\quad\quad\quad  $\mathrm{pC}$ \\
 Bunch length &\quad\quad\quad\quad\quad  20 &\quad\quad\quad\quad\quad  $\mathrm{\mu m}$ \\
 Energy &\quad\quad\quad\quad\quad  $1.5$ &\quad\quad\quad\quad\quad  $\mathrm{GeV}$ \\
 energy spread &\quad\quad\quad\quad\quad  0.06 &\quad\quad\quad\quad\quad $\%$ \\
$\varepsilon_{x_0 \& y_0}$ &\quad\quad\quad\quad\quad 1 &\quad\quad\quad\quad\quad $\mu \mathrm{mrad}$ \\ \hline
 \hline
 \end{tabular}
 \end{table}

\begin{figure} 
  \begin{center}
  \includegraphics[height=5cm]{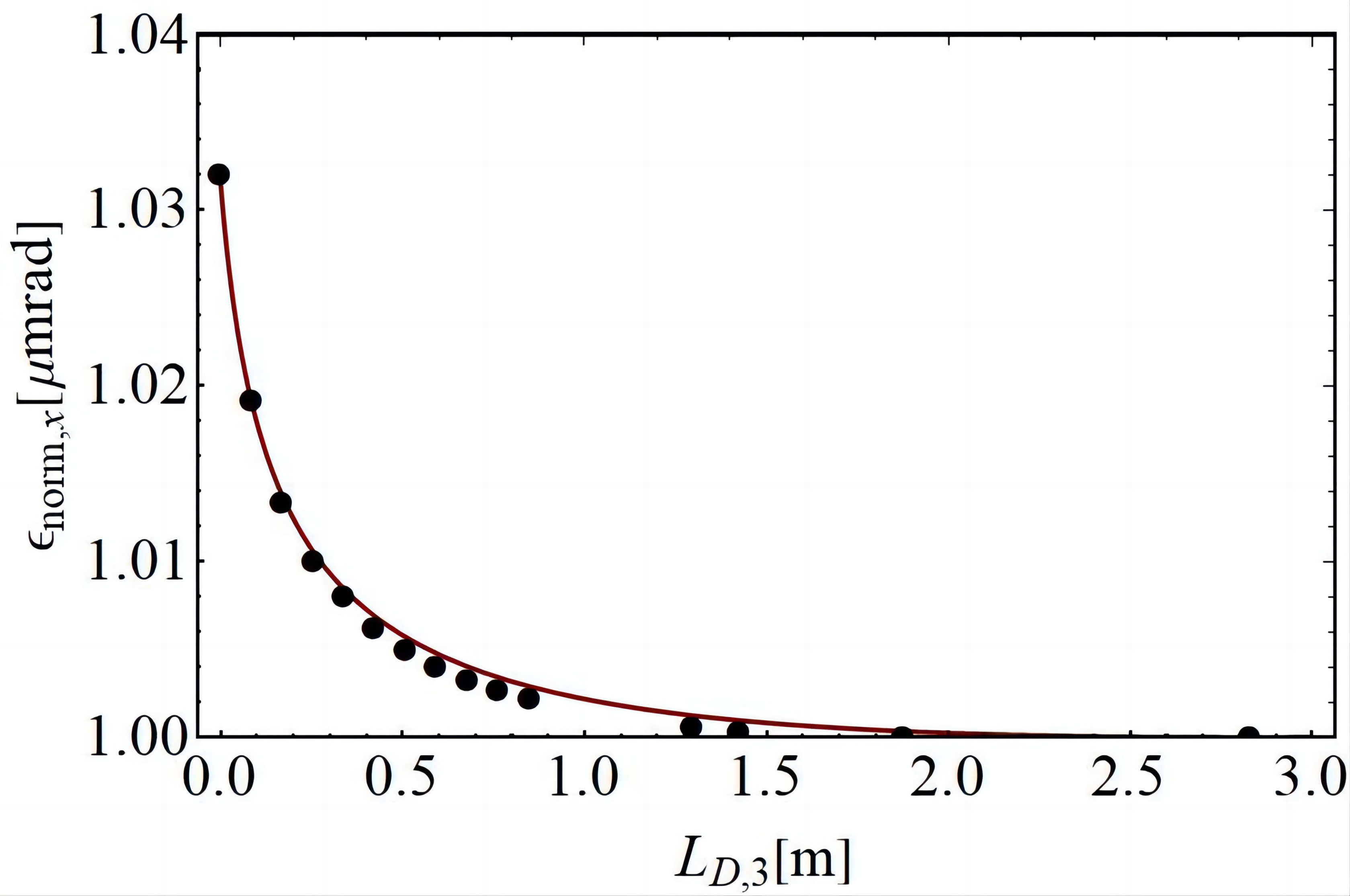}
  \caption{\label{sin_emit}The normalized emittance at the exit of the AQBA with periodic twiss functions. The red line: from Eqs.~(\ref{eq49}). The black dot: from BMAD.}
  \end{center}
  \end{figure}

  Furthermore, CSR in the sequence was taken into consideration. The additional energy spread in this case is related to $s$, as shown in Eqs.~(\ref{eq21}) and (\ref{eq22}). The RMS of the CSR kick cannot be expressed in the simplified form presented in Eqs.(8)-(10) of Ref.\cite{PhysRevAccelBeams.19.064401}. The coupling of the CSR kicks in the two cases is also difficult to solve analytically. In this study, our focus is solely on the cases that the QBA meets the conditions outlined in Eqs.~(\ref{eq8}) As a result, the RMS of the SS CSR kick close to zero, and the coupling between the two types of CSR kicks are negligible. Then the RMS of the CSR kick in $x$ direction can be expressed as
  
 \begin{equation}\label{eq43}
 \begin{aligned}
\left<(x-\left<x\right>)^2\right>&=\left<((x_1-\left<x_1\right>)+(x_2-\left<x_2\right>)+(x_3-\left<x_3\right>))^2\right>\\
x_i&=\frac{4 r_c \lambda(z) R_{52,i} \zeta_i}{\gamma}
 \end{aligned}
 \end{equation}
The RMS of kicks in other directions has a similar form. Then for any $z$ of the bunch, the CSR kicks meet  
  \begin{equation}\label{eq44}
 \begin{aligned}
\frac{x_i(z)}{x_j(z)}&=\frac{R_{52,i}\zeta_i}{R_{52,j}\zeta_j},\\
\frac{x_i(z)}{x'_j(z)}&=-\frac{R_{52,i}\zeta_i}{R_{51,j}\zeta_j}.\\
 \end{aligned}
 \end{equation}
 The correlation coefficient between $x_i$ and $x_j$ is equal to 1. Then the coupling between $x_i$ and $x_j$ can be replaced by
 \begin{equation}\label{eq45}
 \begin{aligned}
\left<(x_i-\left<x_i\right>)(x_j-\left<x_j\right>)\right>&=\sqrt{\left<(x_i-\left<x_i\right>)^2\right>\left<(x_j-\left<x_j\right>)^2\right>}\\ \\
\left<(x_i-\left<x_i\right>)(x'_j-\left<x'_j\right>)\right>&=\sqrt{\left<(x_i-\left<x_i\right>)^2\right>\left<(x'_j-\left<x'_j\right>)^2\right>}.
\end{aligned}
 \end{equation}
 Then the RMS of the CSR kicks in the sequence drifts can be expressed as
  \begin{equation}\label{eq46}
 \begin{aligned}
\left<(x-\left<x\right>)^2\right>=\left(\sqrt{\frac{2\sqrt{3}-3}{3 \pi}}\frac{Nr_c}{\gamma\sigma_z}\right)^2\sum_{i,j}R_{51,i}R_{51,j}\zeta_{i}\zeta_{j}
\end{aligned}
 \end{equation}
   \begin{equation}\label{eq47}
 \begin{aligned}
\left<(x'-\left<x'\right>)^2\right>=\left(\sqrt{\frac{2\sqrt{3}-3}{3 \pi}}\frac{Nr_c}{\gamma\sigma_z}\right)^2\sum_{i,j}R_{52,i}R_{52,j}\zeta_{i}\zeta_{j}
\end{aligned}
 \end{equation}
    \begin{equation}\label{eq48}
 \begin{aligned}
\left<(x-\left<x\right>)(x'-\left<x'\right>)\right>=-\left(\sqrt{\frac{2\sqrt{3}-3}{3 \pi}}\frac{Nr_c}{\gamma\sigma_z}\right)^2\\
\times \sum_{i,j}R_{52,i}R_{51,j}\zeta_{i}\zeta_{j}
\end{aligned}
 \end{equation}
 Similarly, after considering CSR in the sequence drifts, the emittance at the exit of AQBA with periodic Twiss functions is
 \begin{widetext}
     \begin{equation}\label{eq49}
 \begin{aligned}
 \varepsilon_x^2=\varepsilon_{x 0}^2+\varepsilon_{x 0}\left(\sqrt{\frac{2\sqrt{3}-3}{3 \pi}}\frac{Nr_c}{\gamma\sigma_z}\right)^2\left(\sqrt{\frac{M_{12}^2}{1-M_{11}^2}} \sum_{i,j}R_{51,i}R_{51,j}\zeta_i\zeta_j+\sqrt{\frac{1-M_{11}^2}{M_{12}^2}}\sum_{i,j}R_{52,i}R_{52,j}\zeta_i\zeta_j\right)
\end{aligned}
 \end{equation}
 \end{widetext}
A series of simulations were conducted by adjusting the length of $L_3$ in re-lat (specifically, the length of the matching sections between the first and second bends) while accounting for the effects of CSR. It's important to note that $L_{D,3}$ in Fig.~(\ref{sin_emit}) does not represent the actual length of $L_3$, but rather the effective length considering the impact of CSR, similar to replacing $DRIFT$ with $CSRDRIFT$ in ELEGANT\cite{osti_761286}. The simulation results in Fig.~(\ref{sin_emit}) are highly consistent with Eqs.~(\ref{eq49}). The growth in emittance was controlled within $0.02\%$ after considering all CSR cases in the AQBA that meet the conditions stated in Eqs .(\ref{eq20})-(\ref{eq21}). Additionally, the variance in bunch length arc can also be neglected (data not shown). 

  \subsection{simulations for multi-AQBA cells}
The simulations for a single AQBA cell that meets the conditions in Eqs.~(\ref{eq20})-(\ref{eq21}) and Eqs.~(\ref{eq6})-(\ref{eq61}) were discussed in detail in the previous section. It exhibited an excellent ability to suppress CSR kicks. For the case where n isochronous structures are connected in series, $I_1$ and $I_2$ for the $n$-th cell of re-lat can be expressed as
\begin{widetext}
     \begin{equation}\label{eq50}
 \begin{aligned}
\sum_{n=0}^NI_{1,n}&=\frac{\sin(\frac{(N+1)\Phi}{2})\cos(\frac{n\Phi}{2})}{\sin(\frac{\Phi}{2})}I_{1,0}+\sqrt{\frac{M_{12}^2}{1-M_{11}^2}}\frac{\sin(\frac{(N+1)\Phi}{2})\sin(\frac{n\Phi}{2})}{\sin(\frac{\Phi}{2})}I_{2,0},\\ \\
\sum_{n=0}^NI_{2,n}&=-\sqrt{\frac{1-M_{11}^2}{M_{12}^2}}\frac{\sin(\frac{(N+1)\Phi}{2})\sin(\frac{n\Phi}{2})}{\sin(\frac{\Phi}{2})}I_{1,0}+\frac{\sin(\frac{(N+1)\Phi}{2})\cos(\frac{n\Phi}{2})}{\sin(\frac{\Phi}{2})}I_{2,0}\\ \\
\sum_{n=0}^{N}\sum_{i_n}R_{51,i_n}\zeta_{i_n}&=\frac{\sin(\frac{(N+1)\Phi}{2})\cos(\frac{n\Phi}{2})}{\sin(\frac{\Phi}{2})}\sum_{i_0}R_{51,i_0}\zeta_{i_0}+\sqrt{\frac{M_{12}^2}{1-M_{11}^2}}\frac{\sin(\frac{(N+1)\Phi}{2})\sin(\frac{n\Phi}{2})}{\sin(\frac{\Phi}{2})}\sum_{i_0}R_{52,i_0}\zeta_{i,0}\\ \\
\sum_{n=0}^{N}\sum_{i_n}R_{51,i_n}\zeta_{i_n}&=\sqrt{\frac{1-M_{11}^2}{M_{12}^2}}\frac{\sin(\frac{(N+1)\Phi}{2})\sin(\frac{n\Phi}{2})}{\sin(\frac{\Phi}{2})}\sum_{i_0}R_{51,i_0}\zeta_{i_0}+\frac{\sin(\frac{(N+1)\Phi}{2})\cos(\frac{n\Phi}{2})}{\sin(\frac{\Phi}{2})}\sum_{i_0}R_{52,i_0}\zeta_{i,0}
\end{aligned}
 \end{equation}
 \end{widetext}
$\Phi=\arccos((M_{11}+M_{22})/2)$ is the phase advance of the single cell. $I_{i,0}$ represents the integration concerning $R_{5i}$ of the individual cells, $\sum_{i_0}$ is the summarize terms in Eqs.~(\ref{eq16})-(\ref{eq17}) of single AQBA. For single deflection structure, the growth in emittance can be minimized by making SS CSR kicks close to 0 in a certain direction such as the TBA structures in Section.\ref{SA1} and Section.\ref{SD}. However, when the structures are connected in series, the presence of a nonzero $I_{2,1}$ or $I_{1,1}$ term can have an impact on both $I_1$ and $I_2$. Consequently, this can result in both $I_1$ and $I_2$ being nonzero simultaneously for the composite structure. Similar issues arise with CSR kicks within the sequence drifts. As discussed in Section.\ref{SD}, the CSR kicks experienced within the drifts cannot be simultaneously minimized to zero in both directions. Therefore, it is impossible to suppress the CSR-induced growth in emittance after concatenating multiple cells without specially designed matching sections. Since both $I_{1,1}$ and $I_{2,1}$ tend to 0 in this study, the SS CSR response remains close to 0 even after the AQBA is connected in series. Otherwise, the CSR-induced emittance growth may become unpredictable due to the subsequent beamline.

  \begin{figure} 
  \begin{center}
  \includegraphics[height=5cm]{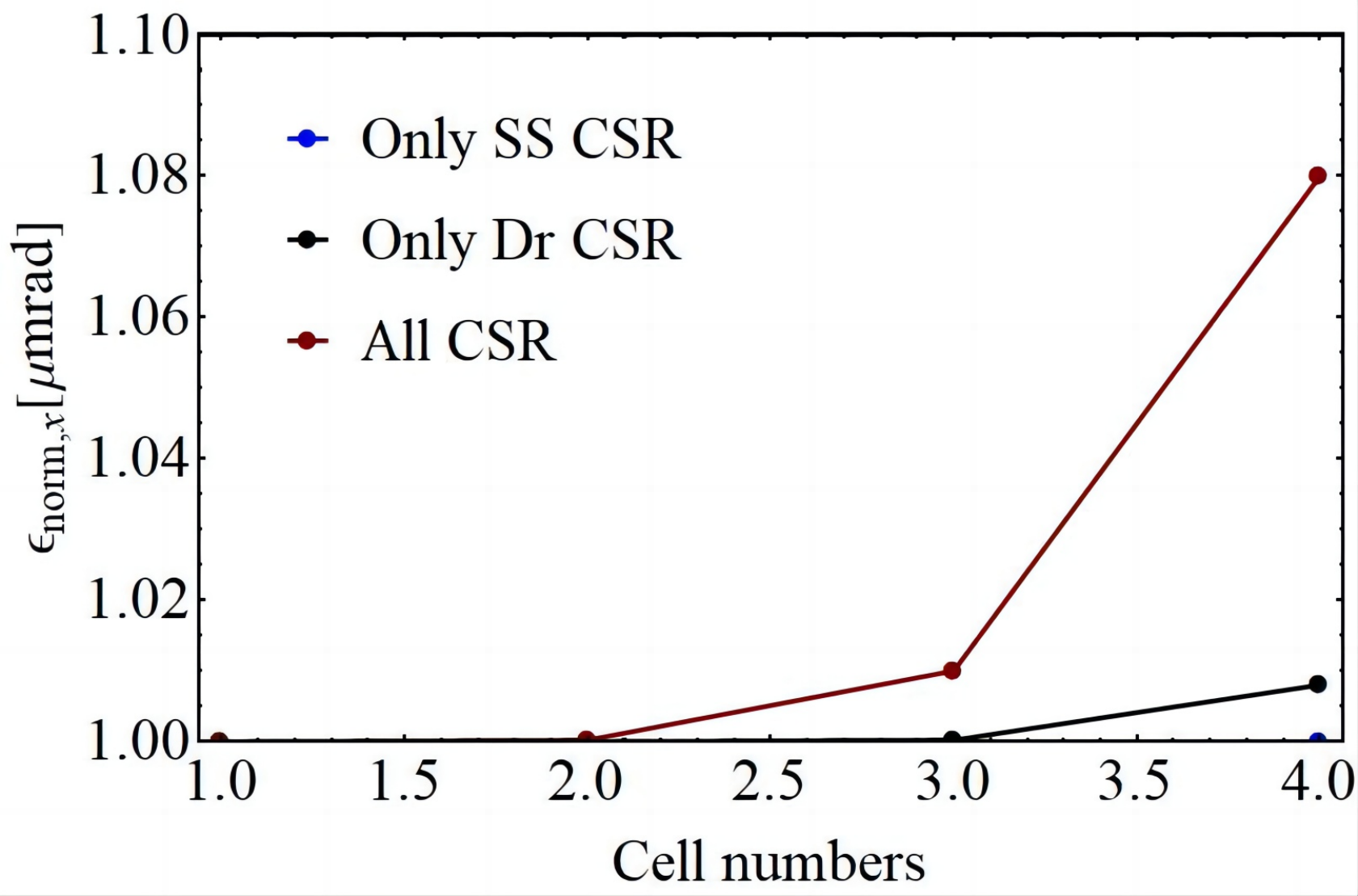}
  \caption{\label{mulit_sim}The normalized emittance at the exit of the AQBA that are connected in series. Three different cases were considered.}
  \end{center}
  \end{figure}

  \begin{table}[!htb]
 \centering
 \caption{Parameters of the non-ideal beams at the entrance of AQBA.}
 \label{beam_tab}
 \begin{tabular}{lll}
 \hline\hline Parameter &\quad\quad\quad\quad\quad  Value &\quad\quad\quad\quad\quad  Units \\
 \hline Bunch charge &\quad\quad\quad\quad\quad  500 &\quad\quad\quad\quad\quad  $\mathrm{pC}$ \\
 Bunch length &\quad\quad\quad\quad\quad  56 &\quad\quad\quad\quad\quad  $\mathrm{\mu m}$ \\
 Energy &\quad\quad\quad\quad\quad  $1.5$ &\quad\quad\quad\quad\quad  $\mathrm{GeV}$ \\
 energy spread &\quad\quad\quad\quad\quad  0.06 &\quad\quad\quad\quad\quad $\%$ \\
$\varepsilon_{x_0 \& y_0}$ &\quad\quad\quad\quad\quad 1.1 &\quad\quad\quad\quad\quad $\mu \mathrm{mrad}$ \\ \hline
 \hline
 \end{tabular}
 \end{table}

  \begin{figure} 
  \begin{center}
  \includegraphics[height=5cm]{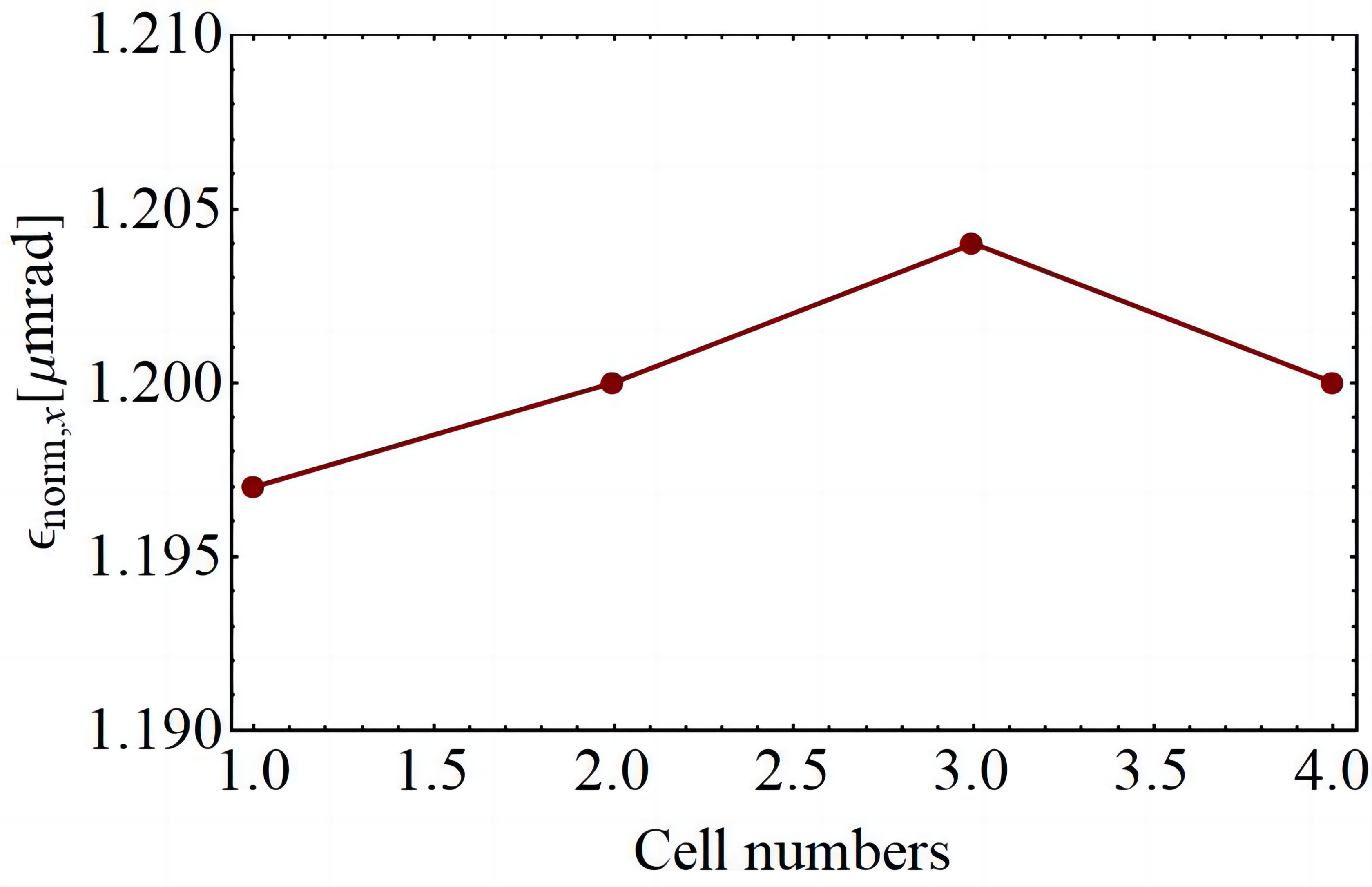}
  \caption{\label{sx}The normalized emittance at the exit of the AQBA beamline for SXFEL bunch. All types of CSR were considered.}
  \end{center}
  \end{figure}

The simulation results for the AQBA structures that are connected in series were presented in Fig.~(\ref{mulit_sim}). Although the derivation in the previous sections was based on Gaussian bunches, the conditions presented in Eqs.~(\ref{eq6})-(\ref{eq61}) and Eqs.~(\ref{eq20})-(\ref{eq21}) were irrelevant to the longitudinal distribution of the bunches. Therefore, we think that AQBA has the capability to mitigate the growth in emittance caused by CSR for any types of bunches. Then, a bunch with a non-ideal distribution was introduced. This non-ideal bunch comes from the Shanghai Soft X-ray Free-Electron Laser (SXFEL)\cite{zhao2011shanghai, zhao2017status}, which is the first X-ray FEL facility in China. The simulation results of SXFEL bunch was shown in Fig.~(\ref{sx}). The variance in emittance can be neglected. For the Gaussian bunch listed in Table.\ref{beam_tab}, the variance in emittance can be neglected when only considering SS CSR or CSR in sequence drifts. When both were considered simultaneously, the growth in emittance was to $8\%$ after passing through four AQBA structures. This is due to the fact that, without the small-angle approximation in bends, the dispersion and $R_{56}$ for the AQBA are not precisely zero. With the accumulation of energy spread cased by CSR, the extra growth in emittance and the variance in bunch length resulting from these errors become increasingly significant and cannot be ignored. A good solution is to further optimize the beamline, disregarding the small-angle approximation in bends. Simultaneously, by reducing the deflection angle of individual bends, local $R_{56}$ can be effectively decreased, resulting in a better preservation of the longitudinal profile. Although the variance in bunch length remains within $1\%$ after connecting four AQBAs in series, decreasing local $R_{56}$ is still beneficial for preserving the longitudinal profile and preventing micro-bunching caused by CSR\cite{PhysRevAccelBeams.20.024401}.

\section{Discussion and Conclusion}
In this study, a novel design approach is introduced. Due to the optimization of re-lat, further derivations for $R_{i6}^{s\rightarrow s_f}$ are rendered unnecessary, which simplifies the optimization process in the design work of more complex structures. Additionally, we propose a QBA capable of fully mitigating SS CSR. It not only possesses periodic Twiss functions but also provides various topologies suitable for different scenarios. Furthermore, the impact of CSR in the sequence drift for ultra-short bunches has been taken into account. A universal condition has been proposed to eliminate emittance growth cased by CSR in the sequence drift, which has subsequently been applied in the optimization of the QBA. Meanwhile, the necessity of QBA for suppressing two types CSR kicks has also been discussed. Simulation results indicate that AQBA demonstrates impressive suppression capabilities in mitigating CSR effects. In future work, the variance in bunch length will be taken into account. Additionally, the related theories are also applicable to the degradation caused by other collective effects, such as longitudinal space charge-induced emittance growth in the deflecting beamline. The related work is currently in progress.

\begin{acknowledgments}
hhhhh
\end{acknowledgments}

\appendix

\section{Analytical Solution for Isochronous TBA Cells with Complete Suppression of SS CSR Effects}

\label{SA1}

 In the case of symmetric TBA configurations with three identical bends, there exists no structure that can simultaneously satisfy all the specified conditions \cite{PhysRevAccelBeams.19.064401,PhysRevAccelBeams.25.090701}. This is evident because the number of constraints exceeds the number of degrees of freedom in this case. Then an additional degree of freedom has been introduced in the form of the ratio between the middle bend and the side bends in the symmetric TBA structure. In the derivation, we assume that all bends have the same radius. Firstly, the structure satisfies the following conditions: 1) $R_{26,\text{mid}}=0$, 2) $R_{56,\text{mid}}=0$, 3) $\det(\mathbf{M_q})=1$. Using the small angle approximation for the bends, the $I_1$ and $I_2$ for the symmetric TBA cells are
\begin{widetext}
\begin{equation}\tag{A.1}\label{a1}
    \begin{aligned}
\frac{36I_1k^2}{\theta_1^2\rho}&=(k+2 m_{22})\left(28 k+12 k^2+8 k^4+3 k^5+8 m_{22}+4 k^3 m_{22}\right)\quad 
\end{aligned}
\end{equation}
\begin{equation}\tag{A.2}\label{a2}
    \begin{aligned}
\frac{36I_2k^2}{\theta_1^3\rho^2}&=(k+m_{22})\left(28 k+12 k^2+8 k^4+3 k^5+8 m_{22}+4 k^3 m_{22}\right)\quad 
\end{aligned}
\end{equation}
\end{widetext}

\begin{figure} 
\renewcommand{\thefigure}{A.1}
\begin{center}
\includegraphics[height=5cm]{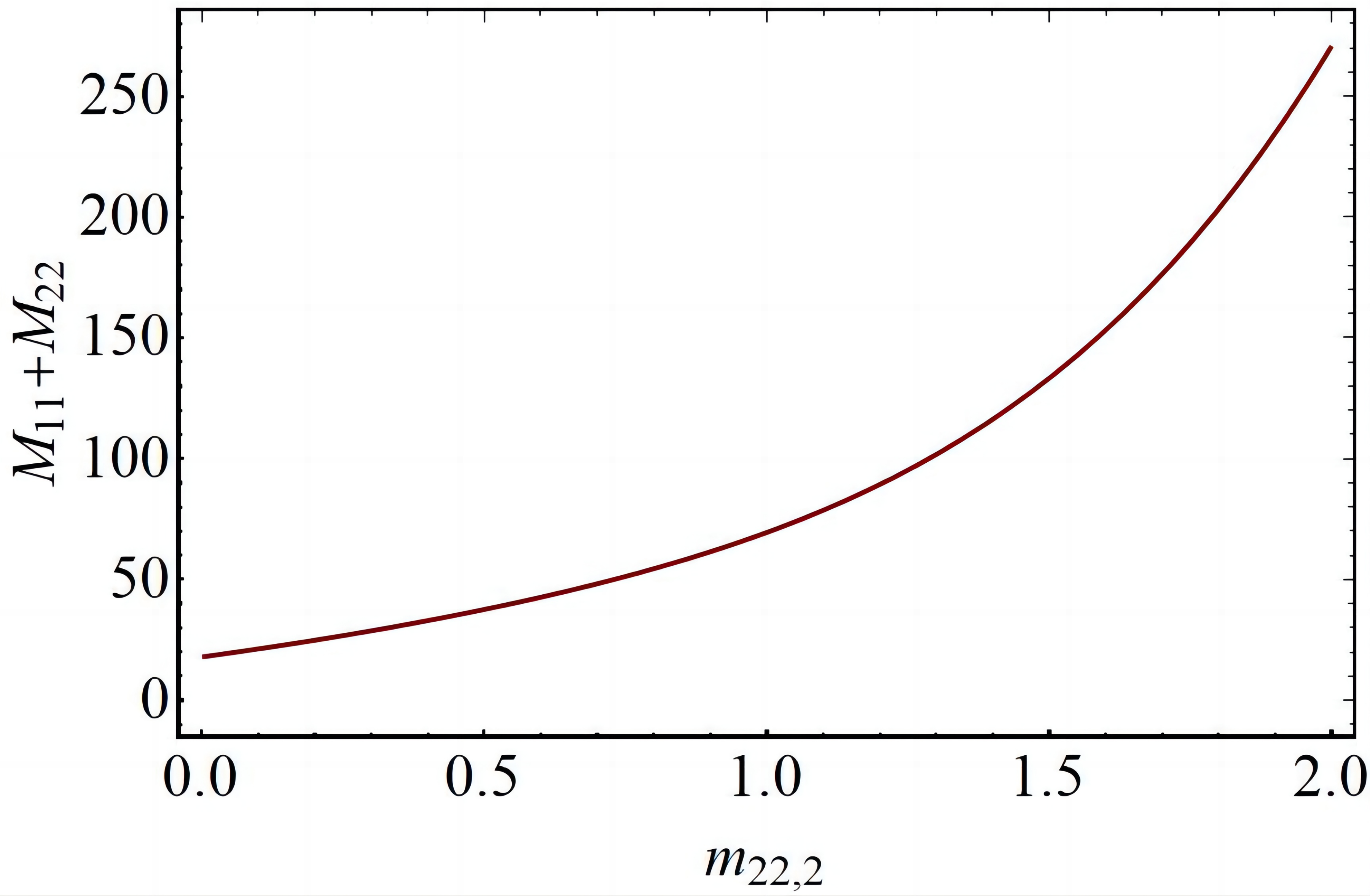}
\caption{\label{fa1} The relationship between the transverse focusing parameter $\left|{M}_{11}+{M_{22}}\right|$ and k, where $k=0$ is a singularity. }
\end{center}
\end{figure}

$\theta_1$ is the angle of the side bends of symmetric TBA cells. $k$ is the ratio between the middle bend and the side bends. $m_{ij}$ are the elements of the transfer matrix from the first bend to the second bend. To ensure that either $I_1$ or $I_2$ equals zero, the value of $m_{22}$ must satisfy the following conditions: 

 \begin{equation}\tag{A.3}\label{a3}
 \begin{aligned}
 I_1=0\rightarrow \begin{cases} m_{22}=-\frac{k}{2} \\ \\ m_{22}=-\frac{28 k+12k^2+8k^4+3k^5}{4(2+k^3)}.\end{cases}
 \end{aligned}
 \end{equation}
 \begin{equation}\tag{A.4}\label{a4}
 \begin{aligned}
 I_2=0\rightarrow \begin{cases} m_{22}=-k \\ \\ m_{22}=-\frac{28 k+12k^2+8k^4+3k^5}{4(2+k^3)}.\end{cases}
 \end{aligned}
 \end{equation}
The second condition in Eqs.~(\ref{a3}) and Eqs.~(\ref{a4}) share the same form. Therefore, both $I_1$ and $I_2$ will equal zero when $m_{22}$ meets this specific condition. This solution has been extensively discussed when $k=1$ in Ref.\cite{PhysRevAccelBeams.19.064401}. However, the TBA cells that meet this condition exhibit over-defocusing in the $x$-plane, which makes it impossible for this structure to be connected in series to achieve large deflection angles. As shown in Fig.~(\ref{fa1}), although the over-defocusing of TBA cells can be mitigated by reducing the value of k, the value of $M_{11}+M_{22}$ is still far greater than 2. It seems impossible to obtain a periodic solution for the TBA in this case by varying the value of $k$. Then, the other solutions for Eqs.~(\ref{a3}) and Eqs.~(\ref{a4}) are given attention. Although these solutions cannot make CSR kicks simultaneously zero in both the $x$ and $xp$ directions, the growth in emittance caused by SS CSR can be controlled according to Eqs.~(\ref{eq41}). The values of $M_{11}+M_{22}$ for these respective cases are 2 ($m_{22}=-k$) and -2 ($m_{22}=-k/2$). Although there are no periodic solutions for the beamline in these specific cases, the asymptotic solution in the vicinity of these scenarios presents a viable solution approach. Related work can be found in Ref\cite{PhysRevAccelBeams.24.060701}.

The asymmetric TBA with three identical bends was considered. And the elements of the transfer matrices for the matching sections between the first and second bends of the re-lat were represented as $m_{ij,1}$. And the for the second matching sections of the re-lat were denoted as $m_{ij,2}$. $M$ was the whole TBA cells. There are six degrees of freedom and six constraints for the asymmetric TBA cells. From conditions 1)-4) in Section.III, there are two sets of solutions:

Case A:
\begin{equation}\tag{B.1}\label{as1}
\begin{aligned}
m_{11,1}=&-5(5+2 \sqrt{8-m _{22,2}})+m_{22,2}\\ \\
m_{12,1}=&\frac{1}{2} L_B(27+11 \sqrt{8-m_{22,2}}-m_{22,2})\\ \\
m_{21,1}=&\frac{66+22 \sqrt{8-m_{22,2}}-2 m_{22,2}}{L_B}\\ \\
m_{22,1}=&-12(3+ \sqrt{8-m_{22,2}})+m_{22,2}\\ \\
m_{11,2}=&-1+ 2\sqrt{8-m_{22,2}}+m_{22,2}\\ \\
m_{12,2}=&-\frac{1}{2}L_B(3+ \sqrt{8-m_{22,2}}+m_{22,2})\\ \\
m_{21,2}=&\frac{2(-3+ \sqrt{8-m_{22,2}}+m_{22,2})}{L_B}.    
\end{aligned}
\end{equation}

\begin{figure} 
\renewcommand{\thefigure}{B.1}
\begin{center}
\includegraphics[height=4cm]{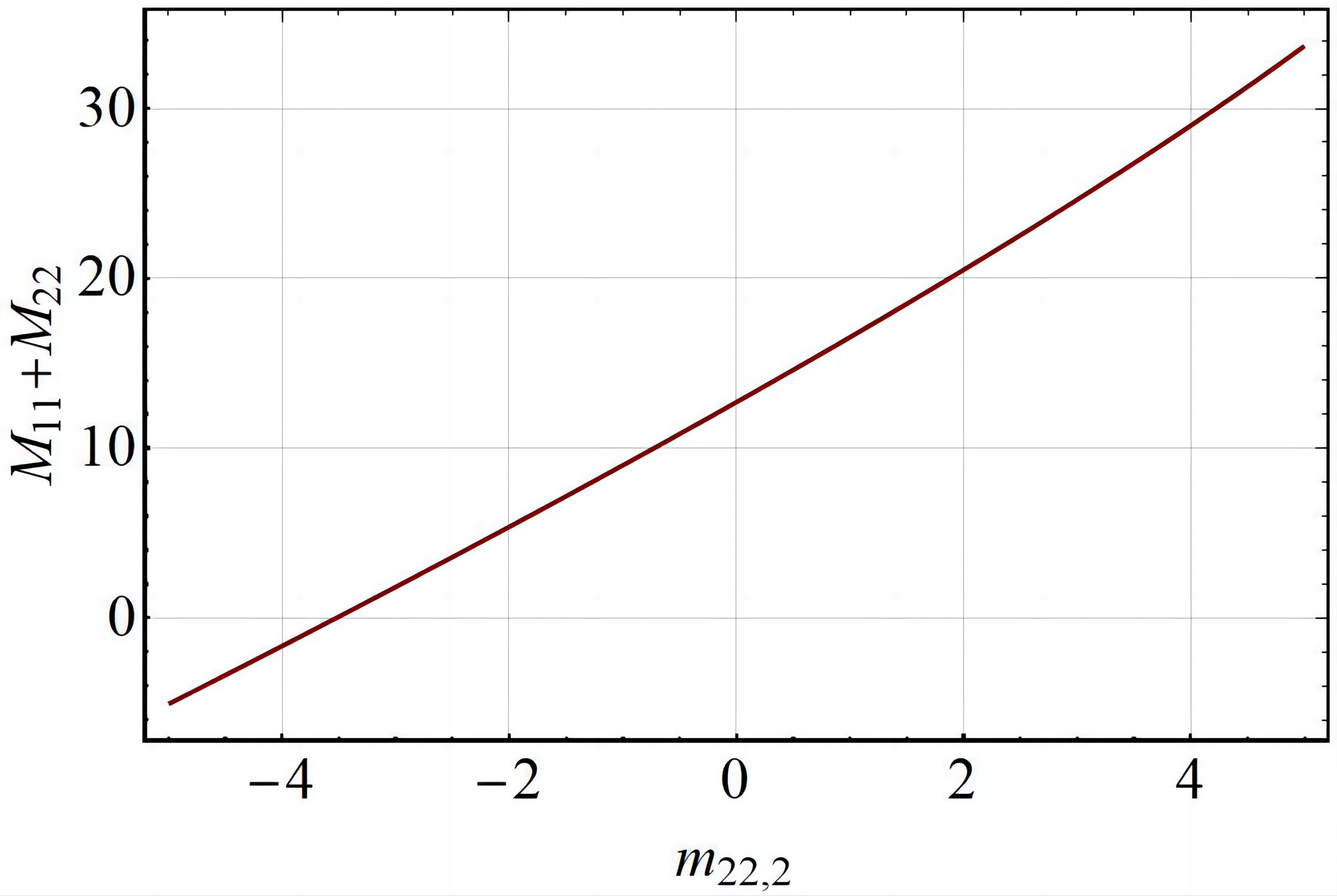}\\ \includegraphics[height=4cm]{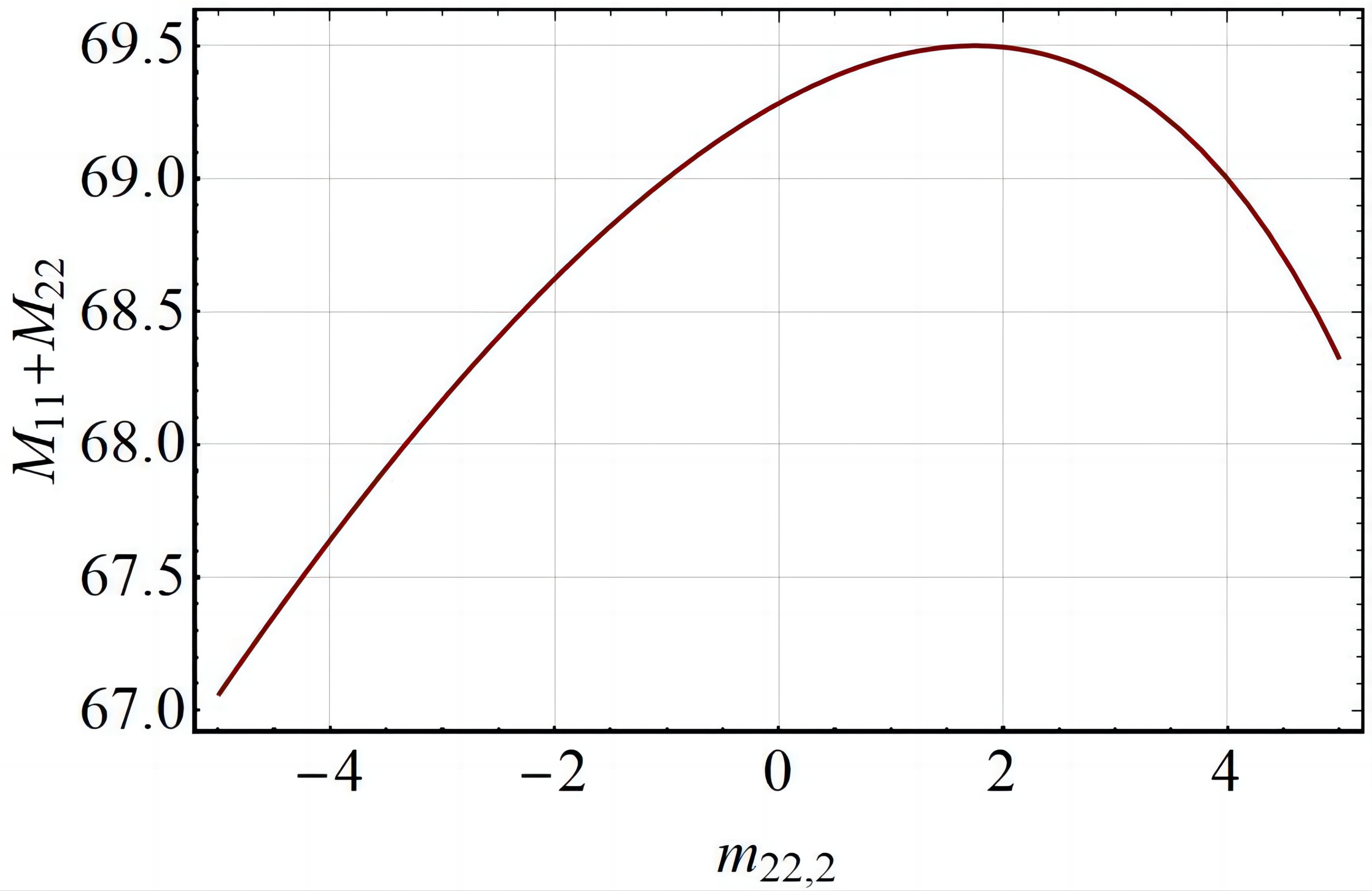}
\caption{\label{fb1} The relationship between $\left|M_{11}+M_{22}\right|$ and $m_{22,2}$. (top) the case A, (bottom) the case B.}
\end{center}
\end{figure}

\begin{figure} 
\renewcommand{\thefigure}{B.2}
\begin{center}
\includegraphics[height=4cm]{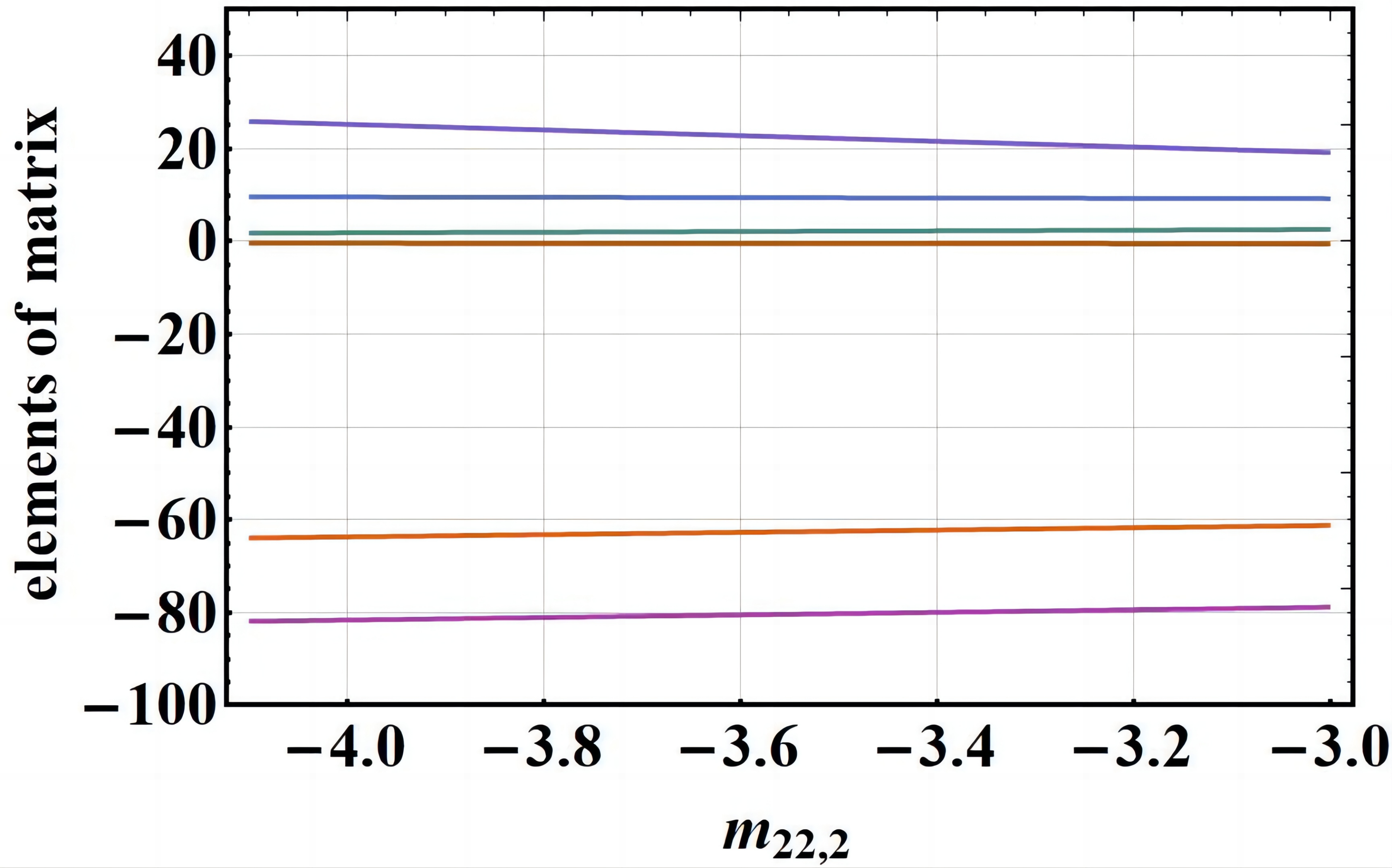}
\caption{\label{fb2} The value of other elements in TBA structures. The curves with each different color represents different elements of the transfer matrix.}
\end{center}
\end{figure}

Case B:
\begin{equation}\tag{B.2}\label{as2}
\begin{aligned}
m_{11,1}=&-5(5-2 \sqrt{8-m _{22,2}})+m_{22,2}\\ \\
m_{12,1}=&\frac{1}{2} L_B(27-11 \sqrt{8-m_{22,2}}+m_{22,2})\\ \\
m_{21,1}=&\frac{66-22 \sqrt{8-m_{22,2}}-2 m_{22,2}}{L_B}\\ \\
m_{22,1}=&12(-3+ \sqrt{8-m_{22,2}})+m_{22,2}\\ \\
m_{11,2}=&-1- 2\sqrt{8-m_{22,2}}+m_{22,2}\\ \\
m_{12,2}=&-\frac{1}{2}L_B(3- \sqrt{8-m_{22,2}}+m_{22,2})\\ \\
m_{21,2}=&\frac{2(3+ \sqrt{8-m_{22,2}}-m_{22,2})}{L_B}
\end{aligned}
\end{equation}
$L_B$ was the total length of the bends. Then the $L_B$ was set to 0.35m (the angle of the bends: $5^{\circ}$, the radii: 4m). Furthermore, the $|M_{11}+M_{22}|$ of the two TBA cases can be altered by adjusting the dimensionless quantity $m_{22,2}$. The range of $m_{22,2}$ is restricted to between -5 and 5 to avoid excessive focusing or defocusing for bunches. The relationship between the $|M_{11}+M_{22}|$ and the $m_{22,2}$ was shown in Fig.~(\ref{fb1}). There are no periodic solutions for the TBA of case B within the specified range.  Additionally, defocusing for bunches is more severe compared to the analytical solution of symmetric TBA. For case A, the TBA exhibits periodic solutions when $m_{22,2}$ is within the range of [-4.1, -3].  However, as shown in Fig.~(\ref{fb2}), the absolute values of other elements in TBA structures are large within this specified range. This could lead to additional beam loss. Therefore, asymmetric isochronous TBA cells with identical bends are not an effective solution for completely canceling the CSR kick.

 \section{The solutions for the four types of QBA cells that meet all the conditions mentioned in Section.\ref{S2} }
\label{SA}

For the chicane-like type:
\begin{equation}
\tag{B.1}
 \begin{aligned}
\label{a1}
m_{11,1}&=\frac{1}{6 \zeta}\left(12+L_B m_{21,1}(4-3 \zeta)\right) \\ \\
m_{12,1}&=\frac{L_B\left(6+m_{22,1}(3 \zeta-4)\right)}{6 \zeta} \\ \\
m_{11,2}&=-\frac{\zeta(3 \zeta-2)\left(5 \zeta-m_{22,1}-3\right)}{m_{22,2}\left(16+3 \zeta+\zeta_{+--+}\right)} \\ \\
m_{12,2}&=\frac{L_B(3 \zeta-2)\left(8+15 \zeta-3 m_{22,1}+\frac{1}{2} \zeta_{+--+}\right)}{6 m_{22,2}\left(16+3 \zeta+\zeta_{+--+}\right)} \\ \\
m_{21,2}&=-\frac{6 m_{22,2}(\zeta-2)}{L_B(3 \zeta-2)} \\ \\
\zeta_{+--+}&=\frac{12-8 m_{22,1}}{2 m_{22,1}+m_{21,1} L_B}\\
 \end{aligned}
\end{equation}

For the zigzag-type:
\begin{equation}\tag{B.2}\label{a2}
 \begin{aligned}
m_{11,1}&=\frac{1}{6 \zeta}\left(12+L_B m_{21,1}(4-3 \zeta)\right) \\ \\
m_{12,1}&=-\frac{L_B\left(6+m_{22,1}(3 \zeta-4)\right)}{6 \zeta} \\ \\
m_{11,2}&=-\frac{(\zeta-2)\left(8-3 m_{22,1}+15 \zeta+\frac{\zeta+-+-}{2}\right)}{m_{22,2}\left(16+3 \zeta+\zeta_{+-+-}\right)} \\ \\
m_{12,2}&=\frac{L_B(3 \zeta-2)\left(8-3 m_{22,1}+15 \zeta+\frac{\zeta_{+-+-}}{2}\right)}{6 m_{22,2}\left(16+3 \zeta+\zeta_{+-+-}\right)} \\ \\
m_{21,2}&=-\frac{6 m_{22,2} \zeta\left(5 \zeta-m_{22,1}-3\right)}{L_B\left(8-3 m_{22,1}+15 \zeta+\frac{\zeta+-+}{2}\right)} \\ \\
\zeta_{+-+-}&=\frac{12-8 m_{22,1}}{2 m_{22,1}+m_{21,1} L_B}
 \end{aligned}
\end{equation}

For the dogleg-type:
\begin{equation}\tag{B.3}\label{a3}
 \begin{aligned}
m_{11,1}&=\frac{1}{6}\left(-4-3 L_B m_{21,1}+\zeta_{++--}\right) \\ \\
m_{12,1}&=-\frac{L_B\left(6+m_{22,1}(4+3 \zeta)\right)}{6 \zeta} \\ \\
m_{11,2}&=-\frac{\zeta(2+3 \zeta)\left(3+5 \zeta-m_{22,1}\right)}{m_{22,2}\left(-16+3 \zeta+\zeta_{++--}\right)} \\ \\
m_{12,2}&=\frac{L_B(2+3 \zeta)\left(-8+15 \zeta-3 m_{22,1}+\frac{1}{2} \zeta_{++--}\right)}{6 m_{22,2}\left(-16+3 \zeta+\zeta_{++--}\right)} \\ \\
m_{21,2}&=-\frac{6 m_{22,2} \zeta\left(5 \zeta+3-m_{22,1}\right)}{L_B\left(15 \zeta-8-3 m_{22,1}+\frac{\zeta++--}{2}\right)} \\ \\
\zeta_{++--}&=\frac{12+8 m_{22,1}}{2 m_{22,1}+m_{21,1} L_B}
 \end{aligned}
\end{equation}
\section{The CSR kick in the sequence drifts in TBA structure. }
\label{SD}
This section continues the discussion from Appendix.\ref{SA1} and solely focuses on scenarios where $k > 0$. For the symmetric TBA cells, the Eqs.~(\ref{eq20}) and Eqs.~(\ref{eq21}) can be simplified as
 \begin{equation}\tag{C.1}
    \begin{aligned}
R_{51,1}&=-R_{51,2} \\
R_{52,1}&=-R_{52,2} \\
\end{aligned}
\end{equation}

For the TBA cells that satisfy Eqs.(A.6), There are:
 \begin{equation}\tag{C.2}
    \begin{aligned}
R_{51,1}&=-\theta_1 \\
R_{51,2}&=-\frac{1}{2} \theta_1\left(6+4 k+2 k^3+k^4\right) \\
R_{52,1}&=-\frac{\theta_1^2 \rho}{2} \\
R_{52,2}&=-\frac{1}{12} \theta_1^2 \rho\left(14+12 k+4 k^3+3 k^4\right)
\end{aligned}
\end{equation}

It is straightforward to infer that there are no solutions in this case. Then we focus on the other cases presented in Appendix.\ref{SA1}. For the cases where $m_{22} = -k/2$ or $m_{22} = -k$, the CSR kick in the $x$ or $x'$ axis disappears. The horizontal variance in emittance can be suppressed by minimizing the $\gamma$ function or $\beta$ function from Eqs.~(\ref{eq41}). Therefore, it is necessary to make the re-lat satisfy Eqs.~(\ref{eq20}) (when $m_{22} = -k/2$) or Eqs.~(\ref{eq21}) (when $m_{22} = -k$) to avoid coupling between CSR in sequence drift and larger Twiss functions. For the case that $m_{22}=-\frac{k}{2}$,
 \begin{equation}\tag{C.3}
\begin{aligned}
& R_{51,1}=-\theta_1 \\
& R_{51,2}=\theta_1 \\
& R_{52,1}=-\frac{\theta_1^2 \rho}{2} \\
& R_{52,2}=\frac{1}{6} \theta_1^2 \rho\left(5+k^3\right).
\end{aligned}
\end{equation}
There are $R_{51,1}=-R_{51,2}$ for any $k$. So it can make the re-lat meet the condition in Eqs.~(\ref{eq20}) for any ratio. However, there is no solutions for $k>0$ that can make the re-lat satisfy $R_{52,1}=-R_{52,2}$. The CSR kicks in the sequence drifts can be minimized by keeping $k$ as small as possible. 

 For the case that $m_{22}=-k$,
  \begin{equation}\tag{C.4}
\begin{aligned}
& R_{51,1}=-\theta_1 \\
& R_{51,2}=\frac{1}{3}(\theta_1-\theta_1 k^3) \\
& R_{52,1}=-\frac{\theta_1^2 \rho}{2} \\
& R_{52,2}=\frac{\theta_1^2 \rho}{2}
\end{aligned}
\end{equation}
 it has $R_{52,1}=-R_{52,2}$ for any value of $k$. Similarly, there are no solutions in this case that satisfy $R_{51,1}=-R_{51,2}$. Additionally, $R_{51}$ remains negative in all regions when $k>1$. 

For the asymmetric case that discussed in Appendix.\ref{SA1}. This structure faces a comparable challenge in symmetric cases where $m_{22}=-\frac{28 k+12k^2+8k^4+3k^5}{4(2+k^3)}$. Because $R_{51}$ and $R_{52}$ remain consistently below zero across the board, eliminating CSR effects during sequence drift becomes impossible. Our only recourse is to minimize the spacing between bends to reduce its influence. 

\section{Derivation of Eqs.~(\ref{eq50})}
In this section, only the first equation in Eqs.~(\ref{eq50}) was considered. Other equations in Eqs.~(\ref{eq50}) have similar forms, thus their derivations were omitted. The first two conditions need to be satisfied initially: 1) the deflection structure is achromatic, and 2) the structure possesses periodic solutions for Twiss functions. For any point within the bend of the $n$-th deflection structure cell ($0$-th represents the first cell in this study). The $R_{5i}$ can be expressed as

  \begin{equation}\tag{D.1}
\begin{aligned}
R_{5i,n}=\left(M^n\right)_{1i}R_{51,0}+\left(M^n\right)_{2i}R_{52,0}
\end{aligned}
\end{equation}

$M$ is the matrix of single cell. Then $M^n$ is the matrix of the beamline consists of a series of n cells. And the integral of $R_{5i}$ of the $n$-th cell is

  \begin{equation}\tag{D.2}
\begin{aligned}
R_{5i,n}=\left(M^n\right)_{1i}I_{1,0}+\left(M^n\right)_{2i}I_{2,0}
\end{aligned}
\end{equation}
$I_{i,0}$ is the integral of $R_{5i}$ of the single cell. Meanwhile, for an achromatic symmetric lattice cell with periodic solution, $M$ can be represented as
  \begin{equation}\tag{D.3}
\begin{aligned}
M=\left(\begin{array}{ll}
\cos\Phi & \beta_{ps}\sin\Phi  \\ \\
\frac{-\sin\Phi}{\beta_{ps}} & \cos\Phi  \\
\end{array}\right)
\end{aligned}
\end{equation}
$\beta_{ps}$ is the periodic solution for beta function. $\Phi$ is the phase advance of single cell. Then the $M^n$ is
  \begin{equation}\tag{D.4}
\begin{aligned}
M^n=\left(\begin{array}{ll}
\cos n\Phi & \beta_{ps}\sin n\Phi  \\ \\
\frac{-\sin n\Phi}{\beta_{ps}} & \cos n\Phi  \\
\end{array}\right)
\end{aligned}
\end{equation}
Furthermore, the integral of $R_{51}$ for the beamline, which is composed of n cells in series, is
 \begin{equation}\tag{D.5}
\begin{aligned}
\label{d5}
I_1=\sum_{n=0}^{N}I_{1,n}=&\left(1+M_{11}+(M^2)_{11}+...+(M^N)_{11}\right)I_{1,0}\\
&+\left(M_{12}+(M^2)_{12}+...+(M^N)_{12}\right)I_{2,0}\\
=&I_{1,0}\sum_{n=0}^{N}\cos(n\Phi)+I_{2,0}\beta_{ps}\sum_{n=0}^{N}\sin(n\Phi)
\end{aligned}
\end{equation}
After expressing Cosine in Complex Form, The frist sum term on the right hand of Eqs.~(\ref{d5}) can be replaced by
\begin{equation}\tag{D.6}
\begin{aligned}
\label{d6}
\sum_{n=0}^{N}\cos(n\Phi)=1+\frac{1}{2} \sum_{n=1}^N\left(e^{i n \Phi}+e^{-i n \Phi}\right)
\end{aligned}
\end{equation}
These sums in the right hand of Eqs.~(\ref{d6}) can be calculated using the geometric series formula:
\begin{equation}\tag{D.7}
\label{d7}
\begin{aligned}
 \sum_{n=1}^N e^{i n \Phi}&=\frac{e^{i \Phi}\left(e^{i N \Phi}-1\right)}{e^{i \Phi}-1}=\frac{e^{i(N+1) \Phi}-e^{i \Phi}}{e^{i \Phi}-1} \\
 \sum_{n=1}^N e^{-i n \Phi}&=\frac{e^{-i \Phi}\left(e^{-i N \Phi}-1\right)}{e^{-i \Phi}-1}=\frac{e^{-i(N+1) \Phi}-e^{-i \Phi}}{e^{-i \Phi}-1}
\end{aligned}
\end{equation}
To simplify, note that the denominators can be written as:
\begin{equation}\tag{D.8}
\label{d8}
\begin{aligned}
 e^{i \Phi}-1&=(\cos\Phi-1)+i \sin\Phi \\
 e^{-i \Phi}-1&=(\cos\Phi-1)-i \sin\Phi
\end{aligned}
\end{equation}
This allows us to rewrite the first sum about cosine as:
\begin{equation}\tag{D.9}
\label{d9}
\begin{aligned}
\sum_{n=0}^{N}\cos n\Phi=&1+\frac{1}{2}\frac{e^{i(N+1)\Phi}-e^{i\Phi}}{(\cos\Phi-1)+i\sin\Phi} \\
&+\frac{1}{2}\frac{e^{-i(N+1)\Phi}-e^{-i\Phi}}{(\cos\Phi-1)-i\sin\Phi}
\end{aligned}
\end{equation}
Using Euler's formulas and simplifying the real parts, we get:
\begin{equation}\tag{D.10}
\label{d10}
\begin{aligned}
\sum_{n=0}^{N}\cos n\Phi=\frac{\sin(\frac{(N+1)\Phi}{2})\cos(\frac{N\Phi}{2})}{\sin(\frac{\Phi}{2})}
\end{aligned}
\end{equation}
Similarly, we can simplified the sum of sine:
\begin{equation}\tag{D.11}
\label{d11}
\begin{aligned}
\sum_{n=0}^{N}\sin n\Phi=\frac{\sin(\frac{(N+1)\Phi}{2})\sin(\frac{N\Phi}{2})}{\sin(\frac{\Phi}{2})}
\end{aligned}
\end{equation}
Furthermore, we can replace $\beta_{ps}$ with $\sqrt{M_{12}^2/(1-M^2_{11})}$ in Eqs.~(\ref{d5}). By substituting the summation term in Eqs.~(\ref{d5}), we obtain the first equation of Eqs.~(\ref{eq50}). Similar derivations can take the other equations of Eqs.~(\ref{eq50}).

%\bibliography{apssamp.bib}% Produces the bibliography via BibTeX.
\providecommand{\noopsort}[1]{}\providecommand{\singleletter}[1]{#1}%

\end{document}